\documentclass[fleqn,usenatbib]{mnras}

\usepackage{newtxtext,newtxmath}
\usepackage[T1]{fontenc}

\DeclareRobustCommand{\VAN}[3]{#2}
\let\VANthebibliography\thebibliography
\def\thebibliography{\DeclareRobustCommand{\VAN}[3]{##3}\VANthebibliography}

\usepackage{graphicx}	% Including figure files
\usepackage{amsmath}	% Advanced maths commands
\usepackage{cuted} % Place in the preamble
\usepackage{pdflscape}
\usepackage{bm}

\newcommand\Dima[1]{{#1}}

\newcommand\Siggi[1]{{#1}}

\title[Long-term orbit integrations with Lobbie]{Fast and Furious: Long-term orbit integrations with collocation integrator Lobbie}

\author[D. Vavilov et al.]{
Dmitrii Vavilov,$^{1,2}$\thanks{E-mail: vavilov@uw.edu; vavilov@iaaras.ru}
Siegfried Eggl,$^{3}$\thanks{E-mail: eggl@illinois.edu}
and Sarah Greenstreet$^{4,1}$\thanks{E-mail: sarah.greenstreet@noirlab.edu}
\\
$^{1}$DiRAC Institute and the Department of Astronomy, University of Washington, 3910 15th Avenue NE, Seattle, WA 98195, USA\\
$^{2}$Institute of Applied Astronomy of the Russian Academy of Sciences, St. Petersburg, Russia \\
$^{3}$The Grainger College of Engineering Department of Aerospace Engineering, University of Illinois Urbana-Champaign, Urbana, IL 61801, USA\\
$^{4}$NSF-DOE Vera C. Rubin Observatory/NSF NOIRLab, 950 N. Cherry Avenue, Tucson, AZ 85719, USA\\
}

\date{Accepted 2026 September 02. Received 2026 July 20; in original form 2026 June 12}

\pubyear{\the\year{}}

\begin{document}
\label{firstpage}
\pagerange{\pageref{firstpage}--\pageref{lastpage}}
\maketitle

\begin{abstract}
%Efficient and accurate numerical integration methods are fundamental to studying long-term solar system dynamics, where numerical errors can accumulate over billions of orbital periods. In this work, we investigate the performance of Lobbie, a Lobatto collocation integrator for long-term simulations of the Solar System. We compare Lobbie against the widely used WHFast symplectic integrator and the high-order adaptive collocation integrator IAS15 implementations in the Rebound package. The performance of the integrator is evaluated using three benchmark problems: a circular two-body orbit integrated over 100 Myrs, the full eight-planet Solar System integrated over 100 Myrs, and the four giant planets integrated over 10 Gyrs. Across all tests, Lobbie demonstrates excellent long-term stability and competitive computational performance. For full Solar System integrations, Lobbie achieves relative energy conservation at the level of $\sim 10^{-14}$ while outperforming IAS15 in computational efficiency and approaching the speed of WHFast at comparable accuracy. Our results demonstrate that high-order Lobatto collocation methods provide a powerful alternative to traditional symplectic mappings or Gauss-Radau collocation methods for long-term dynamical simulations in celestial mechanics.

Efficient and accurate numerical integration methods are fundamental to studying long-term solar system dynamics, where numerical errors can accumulate over billions of orbital periods. In this work, we investigate the performance of \Dima{our modification of} Lobbie, a Lobatto collocation integrator for long-term simulations of the Solar System. We compare Lobbie against the widely used WHFast symplectic integrator and the high-order adaptive collocation integrator IAS15 implementations in the Rebound package. The performance of the integrator is evaluated using \Dima{five} benchmark problems: a circular two-body orbit integrated over 100 Myrs, \Dima{two-body orbits with eccentricities up to $1-10^{-6}$ to test the adaptive step size control during repeated close encounters,} the full eight-planet Solar System integrated over 100 Myrs, the four giant planets integrated over 10 Gyrs, \Dima{and the eight-planet Solar System together with 50 asteroids on Aten-like orbits integrated over 1~Myr to test the integrator's consistency in a chaotic, close-encounter-rich environment.} Across all tests, Lobbie demonstrates excellent long-term stability and competitive computational performance. For full Solar System integrations, Lobbie achieves relative energy conservation at the level of $\sim 10^{-14}$ while outperforming IAS15 in computational efficiency and approaching the speed of WHFast at comparable accuracy. \Dima{In the Solar System plus asteroids test, Lobbie also keeps different realizations of Jupiter's orbit consistent with each other to within about 1.5~km after 1~Myr of integration, two orders of magnitude better than IAS15, being twice faster.} Our results demonstrate that \Dima{Lobbie} provides a powerful \Dima{universal} alternative to traditional symplectic mappings or Gauss-Radau collocation methods for long-term dynamical simulations in celestial mechanics.
\end{abstract}


%%%%%%%%%%%%%%%%%%%%%%%%%%%%%%%%%%%%%%%%%%%%%%%%%%

%%%%%%%%%%%%%%%%% BODY OF PAPER %%%%%%%%%%%%%%%%%%

\section{Introduction}

Understanding the long-term evolution of gravitational systems is one of the central problems in celestial mechanics. Planetary systems, asteroid families, and stellar binaries evolve over timescales that are far longer than direct observations. For this reason, numerical integration of the equations of motion has become a fundamental tool in dynamical astronomy. The reliability of such studies depends strongly on the numerical properties of the chosen integrator, especially when simulations extend over millions or billions of orbital periods.

When integrating $N$-body systems over very long times, two main sources of numerical error must be controlled: truncation error and round-off error. Even very small systematic drifts in conserved quantities, such as the total energy or angular momentum, can accumulate and lead to incorrect physical conclusions. Considerable effort has therefore been devoted to developing numerical schemes that maintain stable behavior over long integrations.

Symplectic integrators have played a central role in this context \citep{Vogelaere1900,1983ITNS...30.2669R,Feng2010,1991AJ....102.1528W}. These methods preserve the geometric structure of Hamiltonian systems and conserve the phase-space properties of the flow. In practice, this often leads to bounded oscillations of the total energy instead of secular drift. Mixed-variable symplectic integrators, in particular, are highly efficient for nearly Keplerian systems, where the motion can be separated into dominant two-body dynamics and smaller perturbations \citep{1991AJ....102.1528W}. For many applications in planetary dynamics, relatively low-order symplectic schemes are sufficient to achieve reliable long-term behavior.

At the same time, symplecticity alone does not guarantee optimal accuracy in all situations. Many modern problems require adaptive timesteps, very high local precision, or the inclusion of additional forces such as relativistic corrections or radiation effects. Moreover, over extremely long integrations, practical implementations may still be affected by round-off accumulation.

\citet{2015MNRAS.446.1424R} demonstrated that very high accuracy and excellent long-term performance can also be achieved with non-symplectic integrators. Their implementation of the \citet{1974CeMec..10...35E}  collocation method, IAS15, combines high order, adaptive timestep control, and careful treatment of round-off errors. Although IAS15 does not preserve the symplectic structure of the equations, it was shown to maintain machine-precision accuracy over extremely long integrations. This result indicates that high-order collocation integrators can provide an alternative path toward reliable long-term simulations in gravitational dynamics.
\Dima{\citet{2018MNRAS.475.5570H} explored this direction further, examining a broader family of collocation methods, based on Gauss-Legendre and Lobatto quadrature, for gravitational $N$-body problems, and showed that such high-order schemes can achieve excellent energy conservation over long integrations, in some cases outperforming standard symplectic integrators.}

In this work, we test the accuracy and efficiency of a relatively new Lobatto type collocation integrator, Lobbie \citep{Avdyushev2021}, for long-term dynamical simulations of the Solar System. Lobbie is a collocation integrator, but it uses Gauss–Lobatto nodes instead of the Gauss–Radau nodes adopted in the Everhart/IAS15 method. The choice of Gauss-Lobatto quadrature imbues the integrator with geometric properties such as time symmetry which can improve integrator performance \citep{hairer2006geometric} \Dima{, though time-symmetry alone does not guarantee long-term energy conservation \citep{2018MNRAS.475.5570H}}.

The aim of this work is to compare Lobbie with the current state of the art algorithms, IAS15 and WHFast, as implemented in the REBOUND package \citep{rein2019high}. We show that Lobbie can produce comparable performance with symplectic algorithms in terms of speed and energy conservation while beating IAS15 in those same metrics. In contrast to symplectic integrators, Lobbie supports arbitrary right hand side function, similar to IAS15. In Section~\ref{sec:integrators} we provide the fundamental ideas behind each of the  integrators, in Section~\ref{sec:tests} we present the results of our tests, and we summarize our findings in the Conclusions.

\section{Numerical integration of the Equations of Motion}
\label{sec:integrators}

Numerical integrators are essential tools in celestial mechanics, enabling the solution of the $N$-body equations of motion, which in general terms read
\begin{equation}{\ddot{\bm{x}}} = \bm{f}(t, \bm{x})
\label{eq:diff_equation}
\end{equation}
where $\bm{x}$ is the vector of the  positions of particles, and hence $\bm{\dot{x}}$ and $\bm{\ddot{x}}$ are vectors of their velocities and accelerations. We will call the function $\bm{f}$ the ''force'' function.

The dynamical state at time $t_0$ provides initial conditions for positions and velocities
\begin{equation}
    \bm{x}_0 = \bm{x}(t_0); \ \bm{\dot{x}}_0 = \bm{\dot{x}}(t_0),
    \label{eq:initial_conditions_Koshi}
\end{equation}
we are looking for the solution at time $t$ by successively advancing integration time by a step size $\Delta t = h$. 
Analytical solutions, for instance for the gravitational N-body problem, can be restricted to few body systems, slow to converge or difficult to find in practice. It is, therefore, often preferable to approximate solutions to equations of motion numerically. A wealth of numerical integration schemes are available and commonly used in dynamical astronomy \citep{eggl2009introduction, rein2019high}. In the following sections, we pitch three of the currently best performing integrators against each other.

\subsection{WHFast}
Consider a Hamiltonian system with state 
$\mathbf{z} = (\mathbf{q},\mathbf{p}) \in \mathcal{M}$, where $\mathbf{q}(t)$ and $\mathbf{p}(t)$ are the generalized coordinates and momenta, i.e. the coordinates in phase space ($\mathcal{M}$). 
The equations of motion in such systems are typically found via Hamilton's equations
\begin{equation}
\dot{\mathbf{q}}=\nabla_\mathbf{p} H, \quad \dot{\mathbf{p}}=-\nabla_\mathbf{q} H,
\end{equation}
where $H(\mathbf{z})$ is the Hamiltonian.
An equivalent formulation can be found using the Lie derivative (or Liouville operator)
$D \equiv \{\cdot, H\}$,
where $\{\cdot,\cdot\}$ denotes the Poisson bracket. For any function $f : \mathcal{M} \to \mathbb{R}$ is that
sufficiently smooth on $\mathcal{M}$, its time evolution is governed by
\begin{equation}
\frac{d}{dt} f(\mathbf{z}(t)) = D f(\mathbf{z}(t)) = \{f, H\}.
\end{equation}
Formally, this defines a linear differential equation in function space,
whose solution can be written in terms of the exponential map of the
operator $D$ as
\begin{equation}
f(\mathbf{z}(t)) = e^{t D} f(\mathbf{z}(0)).
\end{equation}
The operator exponential is understood via its series expansion,
\begin{equation}
e^{tD} = \sum_{n=0}^{\infty} \frac{t^n}{n!} D^n,
\end{equation}
where repeated application of $D$ corresponds to nested Poisson brackets.
The time evolution of the state vector itself is a special case generated by the flow map
\begin{equation}
\mathbf{z}(t) = e^{t D} \mathbf{z}(0),
\end{equation}
which is a canonical transformation preserving the symplectic structure of the underlying dynamics.
This formalism provides the foundation for operator-splitting methods
such as symplectic integrators, where the exponential of a complicated
operator $D = D_A + D_B$ is approximated by compositions of exponentials
$e^{t D_A}$ and $e^{t D_B}$, exploiting the fact that sub-flows along $A$ or $B$ can
often be solved exactly. \citet{1991AJ....102.1528W} and later \citet{rein2015whfast} realized that the gravitational N-body problem with the Hamiltonian
\begin{align}
H &= \sum_{i=0}^{N-1} \frac{p_i^2}{2 m_i}
\;-\;
\sum_{i=0}^{N-1} \sum_{j=i+1}^{N-1}
\frac{G m_i m_j}{\lvert \mathbf{r}_i - \mathbf{r}_j \rvert},
\label{eq:nbodyham}
\end{align}
permits such a splitting, namely 
% \begin{align}
% H_{\pm} &= \sum_{i=1}^{N-1} \frac{G\, m_i\, M_i}{\lvert \mathbf{r}_i \rvert},
% \end{align}
\begin{align}
H = H_0 + H_{\mathrm{Kepler}} + H_{\mathrm{Interaction}},
\end{align}
where $H_0$ encodes the uniform motion of the center of mass, $H_{\mathrm{Kepler}}$ is integrable as a sum of independent two-body problems and $H_{\mathrm{Interaction}}$ represents the interaction between bodies and is independent of the momenta. The Hamiltonian from equation (\ref{eq:nbodyham}) can, for example, be split as follows,
\begin{align}
H &= 
\underbrace{\frac{p_0^2}{2 m_0}}_{H_0}
+
\underbrace{
\sum_{i=1}^{N-1}
\left(
\frac{p_i^2}{2 m_i}
-
\frac{G m_i M_i}{\lvert \mathbf{r}_i \rvert}
\right)
}_{H_{\mathrm{Kepler}}}
\nonumber \\[0.5em]
&\quad +
\underbrace{
\sum_{i=1}^{N-1} \frac{G m_i M_i}{\lvert \mathbf{r}_i \rvert}
-
\sum_{i=0}^{N-1} \sum_{j=i+1}^{N-1}
\frac{G m_i m_j}{\lvert \mathbf{r}_i - \mathbf{r}_j \rvert}
}_{H_{\mathrm{Interaction}}}.
\end{align}
Note that equations for each part of the split Hamiltonian can be solved analytically if they are considered individually. The standard drift--kick--drift scheme then corresponds to
the symmetric composition of individual solutions in the form
\begin{align}
e^{h D} \approx e^{\frac{h}{2} D_0}e^{\frac{h}{2} D_{\mathrm{Kepler}}}
e^{h D_{\mathrm{Interaction}}}
e^{\frac{h}{2} D_{\mathrm{Kepler}}}e^{\frac{h}{2} D_{\mathrm{Kepler}}}e^{\frac{h}{2} D_0},
\end{align}
where each exponential map advances the equations of motion by a time step of $h/2$ and $h$ under the corresponding Hamiltonian, respectively.

Unfortunately, none-commuting operators cannot be split without consequence. In fact, if the commutator of operators $[A,B] = AB - BA\neq0$ then
\begin{equation}
e^{A} e^{B} = \exp\!\left( A + B + \tfrac{1}{2}[A,B]
+ \tfrac{1}{12}[A,[A,B]] - \tfrac{1}{12}[B,[A,B]] + \cdots \right),
\end{equation}
by the Baker--Campbell--Hausdorff (BCH) expansion, where  $\tfrac{1}{2}[A,B]
+ \tfrac{1}{12}[A,[A,B]] - \tfrac{1}{12}[B,[A,B]] + \cdots$ could be considered the ``error terms" of the splitting scheme. 
It is straight forward to show via a BCH expansion that the Wisdom-Holman scheme exactly integrates a modified Hamiltonian
\begin{align}
\begin{split}
\tilde{H} &= H
+ \frac{h^2}{12}
\Big(
\{H_{\mathrm{Interaction}}, \{H_{\mathrm{Interaction}}, H_0+H_{\mathrm{Kepler}}\}\}\\
&- \tfrac{1}{2}
\{H_{\mathrm{Kepler}}, \{H_0+H_{\mathrm{Kepler}}, H_{\mathrm{Interaction}}\}\}
\Big)
+ \mathcal{O}(h^4),
\end{split}\label{eq:modfiedH}
\end{align}
which is identical to the actual Hamiltonian up to $\mathcal{O}(h^2)$.
Since $H_{\mathrm{Interaction}}$ is typically of order $\epsilon$
(the planet-to-star mass ratio), the leading error scales as
$\mathcal{O}(\epsilon h^2)$, which explains the high accuracy
of the method for weakly interacting systems. The error Hamiltonian
is time-independent, implying that the integrator conserves energy
up to bounded oscillations rather than secular drift. To preserve those favorable properties, however, the step size $h$ must remain fixed during the integration, as changing $h$ alters the modified Hamiltonian in equation (\ref{eq:modfiedH}).

WHFast \citep{rein2015whfast} and WHFast512 \citep{javaheri2023whfast512} improve performance by using Jacobian and Democratic Heliocentric Coordinates \citep{Duncan_1998}, faster evaluation of Stumpff functions, as well as advanced vectorization.
Symplectic correctors \citep{wisdom1996symplectic, laskar2001high}, additional operator compositions
designed to cancel low-order BCH terms, can further reduce errors \citep{rein2019symplectic}.
One downside of Wisdom-Holman based maps is that they only resolve deep close approaches between bodies when the integration step size is chosen accordingly.

\subsection{Everhart / IAS15}

The idea of the \citet{1974CeMec..10...35E} integrator, as for other collocation integrators, is to approximate the solution on a step by a polynomial $\bm{u}(t,\bm{x})$ that satisfies the differential equation at $s$ points $\{t_i\}_{i=1}^s$:

\begin{equation}
    \bm{\ddot{u}}(t_i) = \bm{f}(t_i,\bm{u}(t_i)), \ i=1,... s
\end{equation}

In addition, this polynomial must satisfy the initial conditions~(\ref{eq:initial_conditions_Koshi}). From these constraints, the collocation polynomial is determined, and the solution is then obtained as the value of this polynomial at a given time $t$ (for the end of the step we use $t = t_0 + h$). 

To construct the polynomial $\bm{u}$, we first approximate the force function $\bm{f}$ by a polynomial $\bm{p}$ and then integrate it:

\begin{equation}
    \bm{p}(\tau) \equiv \bm{\ddot{u}}(t_0 + h \tau) .
\end{equation}

\noindent where $\tau$ is a uniform parameter $\frac{t - t_0}{h}$ and is in interval $[0,1]$ for points inside one step.
The polynomial $\bm{p}$ is constrained by the conditions $\bm{p}(c_i) = \bm{f}(t_0 + c_i h, \bm{u}(t_0 + c_i h))$. Edgar \citet{1974CeMec..10...35E} proposed to represent $\bm{p}$ not in the Lagrange form (as in classical collocation Runge--Kutta integrators; \citep{hairer2006geometric}), but in the canonical polynomial form:

\begin{equation}
    \bm{p}(\tau) = \sum \limits_{i=0}^{s-1} \bm{\alpha}_i \tau^i .
\end{equation}

Using approximate values $\bm{u}_i$, one computes $\bm{f}_i = \bm{f}(t_0 + c_i h, \bm{u}_i)$ for $i=1,...,s$, and then determines the coefficients $\bm{\alpha}_i$ of the polynomial $\bm{p}$. From $\bm{p}(\tau)$ one reconstructs the polynomial $\bm{u}(t)$, which gives corrected values $\bm{u}_i$. This process is repeated until convergence is reached. For the exact equations we refer the reader to \citet{1974CeMec..10...35E} and \citet{2015MNRAS.446.1424R}.

To obtain the solution on one integration step $h$, the force function must be evaluated $s \cdot n_i$ times, where $n_i$ is the number of iterations required for convergence. In practice, $n_i$ is small, and two iterations are usually sufficient unless there are strong perturbations, such as close encounters for asteroid dynamics. The polynomial coefficients from the current step are then used as a predictor for the next step: $\bm{f}_i = \bm{p}(1 + c_i)$.

An important aspect is the choice of collocation points $\{c_i\}_{i=1}^{s}$. It is known that the maximal order of a collocation integrator is $2s$, but this is achieved only when the points follow the Gauss--Legendre distribution, where no points coincide with the endpoints ($c_1 \neq 0$, $c_s \neq 1$; \citep{hairer2006geometric}). Gauss--Radau quadrature provides the highest possible order, $2s-1$, among schemes where one endpoint is included ($c_1 = 0$ or $c_s = 1$), and this was the choice made by Everhart.

\citet{2015MNRAS.446.1424R} introduced an important improvement in their implementation of the Everhart integrator (IAS15): a new adaptive timestep selection algorithm. This algorithm is dimensionless, which makes it independent of the choice of units. They also showed that for a 15th-order integrator ($s=8$), an optimal accuracy parameter per step is of order $10^{-7}$ in double precision, although a more conservative default value of $10^{-9}$ is typically used.

\subsection{Lobbie}

\citet{Avdyushev2020} developed a new collocation integrator, Lobbie (Lobatto Best Integrator of Equations). Although the general idea is similar to the Everhart integrator, there are several important differences. \citet{Avdyushev2020} preferred  polynomials $\bm{p}$ in Newton form
\begin{equation}
    \bm{p}(\tau) = \sum \limits_{i=1}^{s} \bm{\alpha}_i \prod \limits_{k=1}^{i-1} (\tau - c_k)
\end{equation}
which lead to simpler equations for the coefficients $\{ \bm{\alpha}_i \}_{i=1}^{s}$. We refer the reader to \citet{Avdyushev2021} for details (also in Appendix~\ref{app:lobbie_full} we present the algorithm for a second order differential equation for Lobbie integrator). Another difference is the use of Gauss--Lobatto spacing for the nodes $\{ c_i\}_{i=1}^{s}$. %As mentioned earlier, the choice of these points is very important.

Collocation methods based on Legendre and Lobatto nodes have important geometric properties: they are symmetric, and the Legendre scheme is also symplectic. In general, Legendre nodes are preferred because they provide a higher order (by two) for the same number of points. However, Lobatto methods are slightly faster in practice. For each integration step with $n_i$ iterations, the number of force evaluations is $n_i (s-1)$ for Lobatto nodes, compared to $n_i s$ for Legendre (and Radau) nodes. In addition, Lobatto nodes include the endpoints of the interval, which simplifies the implementation, improves the predictor, and can reduce the number of iterations $n_i$.

In this work, we use the Fortran implementation of the Lobbie integrator provided by \citet{2022SoSyR..56...32A}, but with two important modifications. First, following \citet{2015MNRAS.446.1424R}, we apply a dimensionless timestep selection. In \citet{Avdyushev2021}, the next step $h_{n+1}$ is computed from the previous step $h_n$ as:

\begin{equation}
    h_{n+1} = r h_n; \ r = \left( \frac{s ||\bm{e}_{tol}||}{h_n ||\bm{\alpha}_s||} \right)^{1/s},
    \label{eq:h_avdyushev}
\end{equation}

% \noindent where $\bm{e}_{tol}$ is the tolerated error (vector) of the velocity assigned by the user and $||.||$ is the sign of $L_2$ norm. To make the step selection dimensionless we need inside the parentheses in Equation~(\ref{eq:h_avdyushev}) multiply by something that has the unit of velocity. Here in this work we 

\noindent where $\bm{e}_{tol}$ is the tolerable velocity error vector and $||\cdot||$ denotes the $L_2$ norm. To make this expression dimensionless, we multiply the numerator by a quantity with units of velocity. In this work, we use the following step-size determination:

\begin{equation}
    h_{n+1} = r h_n; \ r = \max \left\{ \left(  \frac{s e_{tol,j} ||\bm{v}_k||}{h_n a_{s,j}} \right )^{1/s} \right\},
\end{equation}

\noindent where $e_{tol,j}$ and $a_{s,j}$ are the $j$-th components of $\bm{e}_{tol}$ and $\bm{\alpha}_s$, and $||\bm{v}_k||$ is the magnitude of the velocity vector for the corresponding particle. The parameter $\bm{e}_{tol}$ is now dimensionless and represents the relative accuracy of the velocity components per step. As noted by \citet{Avdyushev2021}, the timestep is effectively determined assuming order $s-1$ of the integrator, although the true order is $2s-2$. Since it is not possible to construct an error estimate of the same high order, in practice we determine suitable values of $\bm{e}_{tol}$ empirically in the next section for the 14-th order integrator ($s=8$).

The second modification is minor but important: we use quadruple precision for the state vectors. All computations are still performed in double precision, but positions and velocities are stored in quadruple precision (Fortran type \texttt{real(16)}). 
At each step of integration we find the correction and then add it to the state vector. The correction can be several orders of magnitude smaller than the state vector itself. Therefore keeping the state vector in quadruple precision allows us to save several significant digits that otherwise would have been lost. This mechanism helps to reduce round-off errors. 
On the other hand since only a few operations are done in quadruple precision (updating the state vector only), this does not significantly affect performance.
We have tested Kahan summation \citep{10.1145/363707.363723} for a further increase in integration accuracy, similar to \citep{2015MNRAS.446.1424R}, but saw a significant hit in performance and no notable improvement in precision.

\subsubsection{Close encounters in Lobbie}

One of the applications of the Lobbie integrator is the long-term integration of the Solar System together with large numbers of massless particles, such as asteroids, comets, meteoroids, or dust particles. In such problems, close encounters with planets play an important role and require careful numerical treatment.

Lobbie uses an adaptive timestep and in general handles close encounters naturally, since the timestep automatically decreases when a particle approaches a massive body. However, for additional robustness, we implemented a dedicated close-encounter and collision handling algorithm in our $N$-body code.

At every integration step, the distances between all massless particles and massive bodies are checked. If a particle approaches a planet closer than 0.03~au, the integration is returned to the previous step and restarted with stricter numerical parameters. In particular, the timestep is \Dima{decreased 3 times} and the accuracy parameter is reduced to a much smaller value (in our practical applications we use $e_{tol}=10^{-14}$). This allows the code to resolve the encounter region with much higher precision. \Dima{We emphasize that this 0.03~au threshold is not a  definition of a close encounter, and it is not used to identify or classify one; it is simply a fixed distance, chosen to lie  outside the sphere of influence of every planet in the Solar System, that triggers this precautionary switch to higher precision before the particle actually enters the encounter region.}

As the particle continues to approach the planet, \Dima{now integrated with these stricter settings,} the code monitors whether it enters the planetary sphere of influence. \Dima{Only entering the sphere of influence itself is classified as a close encounter}, and the event is recorded as such. If the particle intersects the physical radius of the planet or the Sun, it is treated as a collision and removed from the integration. We also monitor particles leaving the Solar System. If a particle reaches a heliocentric distance of 100~au and simultaneously has an unbound orbit, it is removed from the simulation and classified as an ejected object.

The Fortran implementation of Lobbie for integrations of massive bodies together with massless particles is publicly available at git repository\footnote{$\mathrm{https://github.com/DJVil/Lobbie\_planet\_integration.git}$}.

\section{Tests}
\label{sec:tests}

\begin{figure}
    %\centering
    \includegraphics[width=0.97\columnwidth]{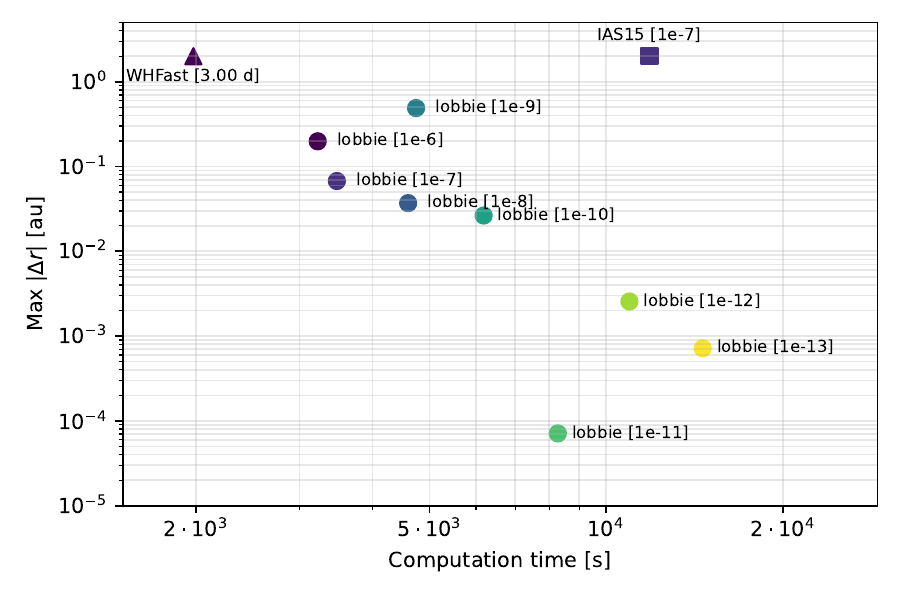} \\
    \includegraphics[width=\columnwidth]{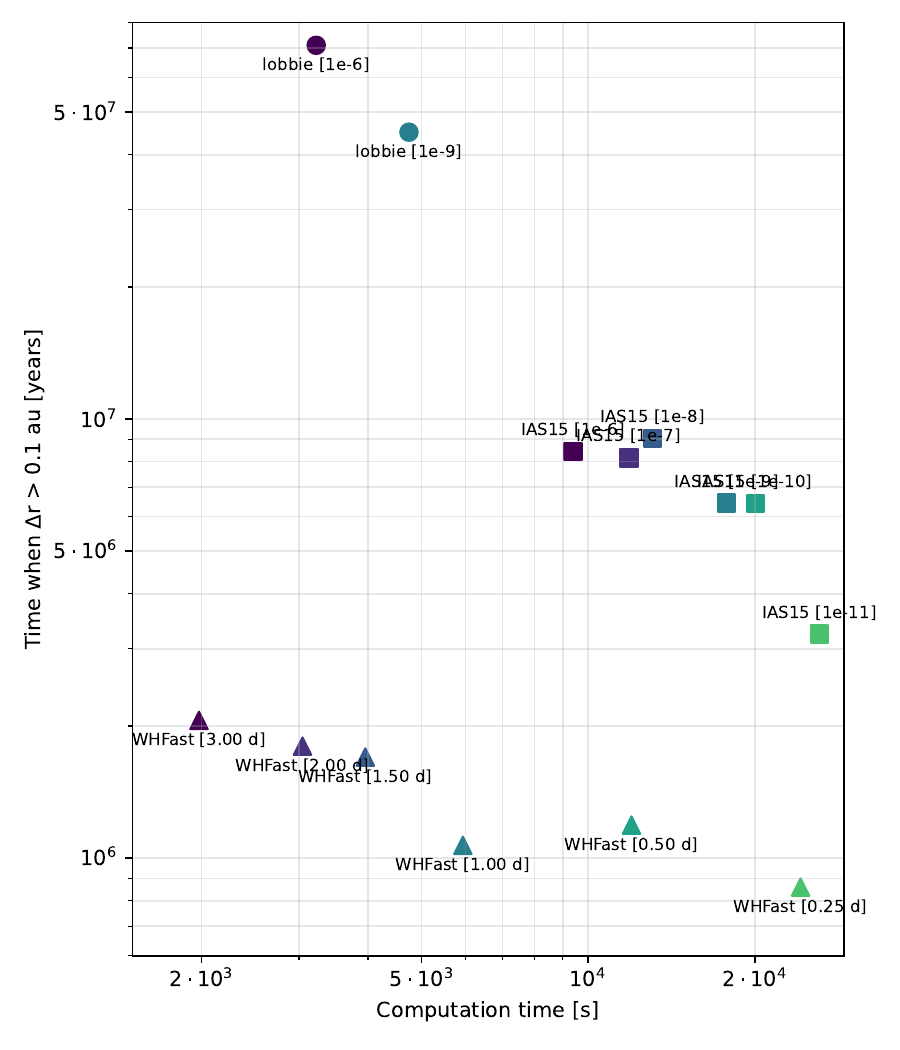}
    \caption{The comparison of Lobbie, IAS15 and WHFast integrators  for the two-body problem. In brackets $[]$ we present the accuracy parameter for Lobbie and IAS15 or the integration timestep in days of WHFast.
    Top: maximum positional error for a massless particle in the two-body problem versus wall clock time it took to run the computations (computation time). All IAS15 and WHFast runs reached 2~au positional differences, hence we show only 1 of the runs for each. Bottom: time at which the positional error reaches 0.1~au versus total computation time (longer is better). Only lobbie[1e-6] and lobbie[1e-9] reached that discrepancy among Lobbie runs. }
    \label{fig:2body}
\end{figure}

To test the efficiency and robustness of the new integrator for long-term dynamics, we performed several numerical experiments and compared the results with WHFast and IAS15. We considered three test scenarios: (a) a two-body problem, 
(b) integration of all eight major planets of the Solar system, and (c) integration of
the four gas giants (Jupiter, Saturn, Uranus, and Neptune).

All integrations were performed in a system of units where the gravitational constant $G=1$ and the mass of the central body is also equal to 1. Distances are measured in astronomical units, therefore a particle with a semi-major axis equal to 1 has an orbital period of $2\pi$. Throughout this work, one year corresponds to $2\pi$ in these units, and we use years when presenting the results.

Each integrator was tested under different accuracy regimes. For Lobbie and IAS15, we varied the dimensionless accuracy parameter $e_{tol}$ in the range $\{10^{-6}, 10^{-7}, 10^{-8}, 10^{-9}, 10^{-10}, 10^{-11}, 10^{-12}, 10^{-13}\}$. The WHFast integrator uses a fixed timestep, and we tested it with timesteps $\{0.25, 0.5, 1, 1.5, 2, 3\}$ days. A timestep of the order of 12 hours is commonly used for Solar System integrations with massless particles such as asteroids \citep{2018Icar..312..181G, 2023AJ....166...55N}. \citet{2012Icar..217..355G} used even a 4~h timestep to properly resolve high-speed encounters with the Earth, Venus and Mercury. Too large timesteps may miss close encounters or even collisions, which leads to a significant loss of accuracy. All tests were performed on AMD EPYC 9654 (96-core, 2400 MHz) processors.

\subsection{Two-body problem}

For the two-body test, we considered a central body of mass 1 and a massless particle on a circular orbit with radius 1. The initial conditions were chosen as $x=\frac{\sqrt{2}}{2}$ and $y=\frac{\sqrt{2}}{2}$. The orbit was integrated for 100 Myr. Every 10,000 years, we recorded the $(x,y)$ position of the particle and compared it with the initial position. Since the orbital period is exactly 1 year, the particle should return to the same position at these times. Any deviation reflects the numerical error of the integrator. \Dima{We note that, in this unperturbed setting, WHFast reduces to a Kepler solver applied repeatedly with a fixed step, since there are no perturbing forces for its symplectic map to integrate.}

Figure~\ref{fig:2body} (top panel) shows the maximum positional error with respect to the analytical solution  as a function of how long it took in seconds to finish the integrations (computation time). After 100 Myr of integration, all WHFast and IAS15 reach errors of  $\sim 2$, meaning that the orbital phase is completely lost. For this reason, we show only the results for IAS15[1e-7] and WHFast[3.00~d]. Instead, the bottom panel shows the time at which the positional error reaches 0.1~au as a function of computation time to reach 100~Myr. The labels indicate the corresponding $e_{tol}$ values (for Lobbie and IAS15) or timesteps (for WHFast). Only lobbie[1e-6] and lobbie[1e-9] among lobbie runs reached the positional error of 0.1~au, that is why the rest are not presented on the bottom panel. However, they kept the precision below 0.1~au for almost 50~Myrs, while all IAS15 and WHFast runs didn't hold it for more than 10~Myrs.

Lobbie demonstrates high robustness. The best result is obtained for $e_{tol}=10^{-11}$, where the positional error remains below $\sim 7 \cdot 10^{-5}$, which is smaller than the diameter of the Earth. At the same time, it is faster than all IAS15 runs and also faster than WHFast with timesteps smaller than 0.5 days. %For WHFast, we observe the expected trend that decreasing the timestep worsens accuracy. 
IAS15 performs better than WHFast, keeping the error below 0.1~au for almost $\sim 10$ Myr for $e_{tol}=10^{-8}$. %, but also shows non-monotonic behavior, with $e_{tol}=10^{-11}$ giving worse results.

\begin{figure}
    %\centering
    \includegraphics[width=1\columnwidth]{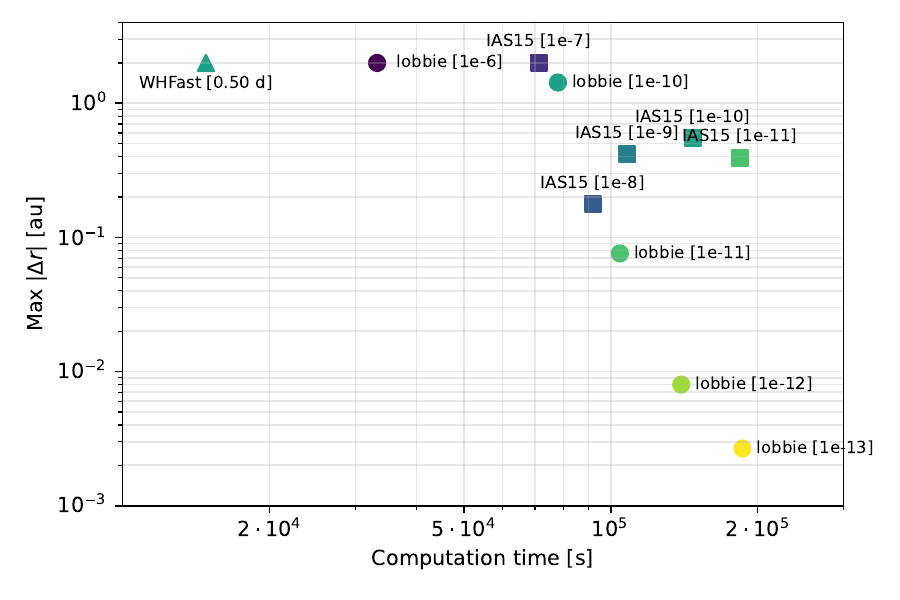} \\
    \includegraphics[width=0.97\columnwidth]{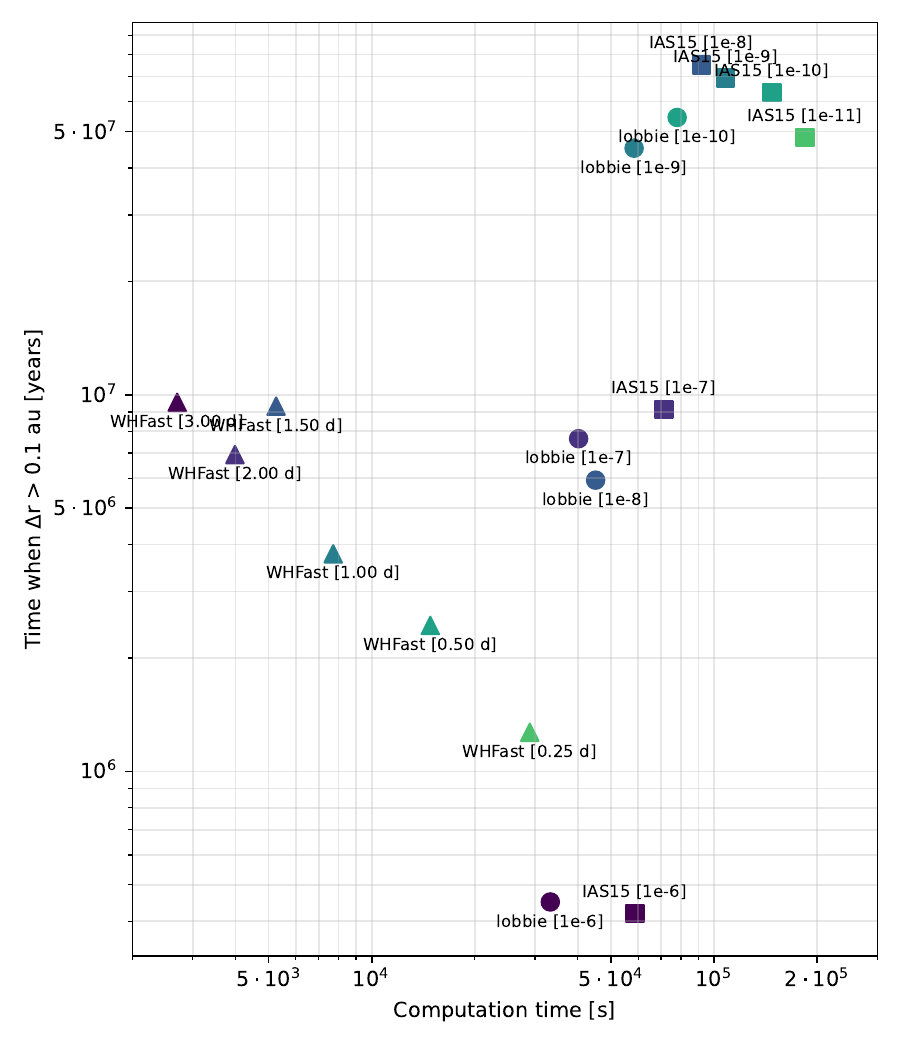}
    \caption{\Dima{Same as Figure~\ref{fig:2body} but for the two-body problem with eccentricity $e=0.999$.
Top: maximum positional error for a massless particle in the two-body problem versus wall clock time needed to run the computation (computation time). As all WHFast runs reached a positional error of 2~au, only one of them is shown. Only Lobbie runs with accuracy parameter smaller than $10^{-10}$ did not reach 2~au maximum positional error. Only IAS15 runs with accuracy parameter smaller than $10^{-8}$ did not reach 2~au positional error. Bottom: time at which the positional error reaches 0.1~au versus total computation time (longer is better). Lobbie runs with accuracy parameter less than $10^{-11}$ did not reach a positional error of 0.1~au, so they are not shown in this panel.}}
    \label{fig:2body_eccentric}
\end{figure}

%\Dima{In order to test Lobbie's adaptive step change we perform a similar two-body test but with an object on an eccentric orbit. We make the test with eccentricity $e=0.999$. The initial conditions are that we put the particle at its aphelion with a distance from the Sun of $Q = 1.999$~au, but keeping the semimajor axis $a=1$~au. The particle's absolute velocity is computed from the integral of energy: $ v^2 = (\frac{2}{Q}-\frac{1}{a}) $. Hence the particle has constantly have close encounters with the Sun at 0.001~au and the step size is being reduced by several orders (for $e_{tol}=10^{-11}$ it is reduced by 5 orders of magnitude from 8.7~days to 7.5~seconds).}

\Dima{To test Lobbie's adaptive step size change, we performed a similar two-body test, but now with the object on an eccentric orbit, with $e=0.999$. We placed the particle at aphelion, at a distance from the Sun of $Q=1.999$~au, while keeping the semi-major axis at $a=1$~au. The particle's velocity was computed from the integral of energy: $v^2 = \left(\frac{2}{Q}-\frac{1}{a}\right)$. With these initial conditions, the particle has a close encounter with the Sun at every orbit, coming as close as 0.001~au. Because of this, the step size is reduced by several orders of magnitude (for $e_{tol}=10^{-11}$, it drops by about 5 orders of magnitude, from 8.7~days down to 7.5~seconds).}

%\Dima{The results are presented on Figure~\ref{fig:2body_eccentric}. First thing we notice is the the computation time increased by an order of magnitude as a consequence of the required stepsize change. As for a circular orbits all WHFast runs reached 2~au positional error after 100~Myr. Surprisingly IAS15 showed a better robustness compared to a circular orbit. With accuracy parameter less than $10^{-8}$ the runs did not reach 2~au position error. We see that the error is minimal for $10^{-8}$ and starts increasing for the other accuracy parameters. This is probably due to the round-off errors.}

\Dima{The results are shown in Figure~\ref{fig:2body_eccentric}. The first thing to notice is that the computation time increased by about an order of magnitude, as a consequence of the much smaller step size required. As in the circular case, all WHFast runs reached a positional error of 2~au after 100~Myr. IAS15 showed better robustness here than in the circular case: runs with accuracy parameter smaller than $10^{-8}$ did not reach 2~au positional error. The best performance occurred at $e_{tol}=10^{-8}$ and increases again for smaller values of $e_{tol}$, which is likely caused by round-off error.}

%\Dima{Lobbie again performed the best compared to IAS15 and WHFast. Starting from $e_{tol}=10^{-10}$ it does not reach 2~au positional error and achieved the best accuracy of $2.67\cdot 10^{-3}$ with $e_{tol}=10^{-13}$. We clearly see that downhill trend with improving accuracy with smaller $e_{tol}$ parameter. This is probably the result of our methodology for mitigating with round-off errors: using quadruple precision for keeping the state vectors. This test is a good verification of the ability of our implementation of Lobbie to work with close encounters. }

\Dima{Lobbie again performed best, compared to both IAS15 and WHFast. Starting from $e_{tol}=10^{-10}$, Lobbie does not reach 2~au positional error, and it achieves its best accuracy of $2.67\cdot 10^{-3}$~au at $e_{tol}=10^{-13}$. There is a clear trend: accuracy keeps improving as $e_{tol}$ decreases, without the round-off-driven turnaround seen for IAS15. We believe this is because of our strategy for handling round-off error, namely storing the state vectors in quadruple precision. }

\Dima{This test is a good demonstration that our implementation of Lobbie can handle close encounters. It is also a particularly useful test case, for two reasons. First, because the two-body problem has an exact analytical solution, we can measure the true error of the integration directly, without relying on energy or angular momentum conservation as a proxy for accuracy. Second, because the orbit is periodic, the particle passes through the same close encounter with the Sun many times during the 100~Myr integration, not just once. This means that the test does not simply check whether an integrator can survive a single close encounter, but whether it can handle a large number of them in a row, while still returning to the correct position each time. In this way, a single test combines a strong check of adaptive step size control with a reliable, analytically known measure of accuracy.}

%\Dima{We provide the third test, which is very similar to the previous one but with $e=1-10^{-6}$. For such an extreme case the distance between the particle and the Sun becomes 1 millionth au, far inside the Sun.  During the close encounter Lobbie with $e_{tol}=10^{-11}$ lowered down the stepsize to $2.5\cdot 10^{-4}$ seconds. All the runs of all integrators reached the 2~au positional error way before the end of the integrations. Therefore on Figure~\ref{fig:2body_super_eccentric} we present only the time when the integrations reached 0.1~au positional error. Again Lobbie with $e_{tol}=10^{-13}$ shows the best results.}

\begin{figure}
    %\centering
    \includegraphics[width=0.97\columnwidth]{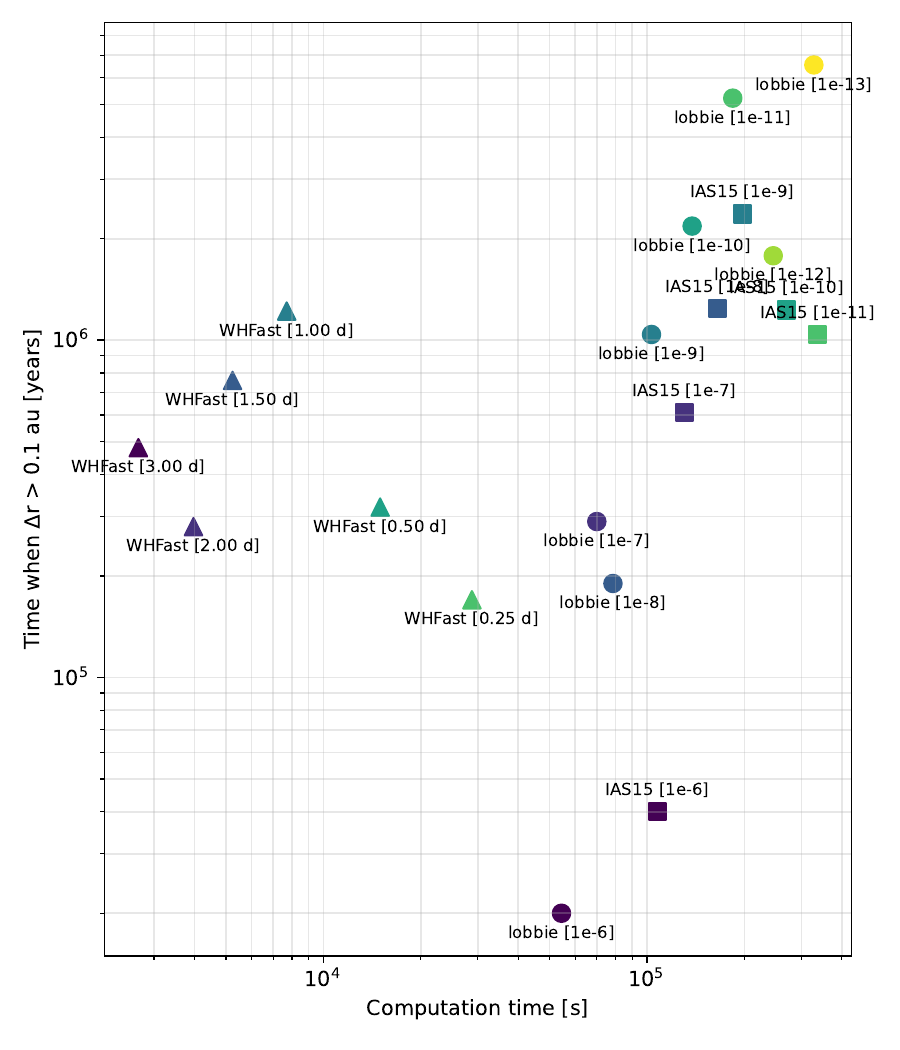}
    \caption{\Dima{Same as Figure~\ref{fig:2body} (bottom) but the comparison for the two-body problem with an eccentricity $e=0.999999$.
    Time at which the positional error reaches 0.1~au versus total computation time (longer is better).  }}
    \label{fig:2body_super_eccentric}
\end{figure}

\Dima{We ran a third test, very similar to the previous one but with $e=1-10^{-6}$. In this extreme case, the minimum distance between the particle and the Sun is about one millionth of an au, well inside the Sun. During the close encounter, Lobbie with $e_{tol}=10^{-11}$ reduced the step size down to $2.5\cdot 10^{-4}$~seconds. All runs, for all integrators, reached a positional error of 2~au well before the end of the integration. For this reason, in Figure~\ref{fig:2body_super_eccentric} we only show the time at which each run reached a positional error of 0.1~au. Once again, Lobbie with $e_{tol}=10^{-13}$ gives the best result.}

\subsection{Integrating the Solar system}

For the Solar System integrations, we used initial positions and velocities from the DE440 planetary ephemeris \citep{2021AJ....161..105P} at epoch Jan 1, 2000 (the exact state vectors are presented in Appendix~\ref{sec:state_vectors_of_planets}). The adopted planetary masses (in units of the solar mass) are listed in Table~\ref{tab:masses}.

\begin{figure}
    \centering
    \includegraphics[width=1\columnwidth]{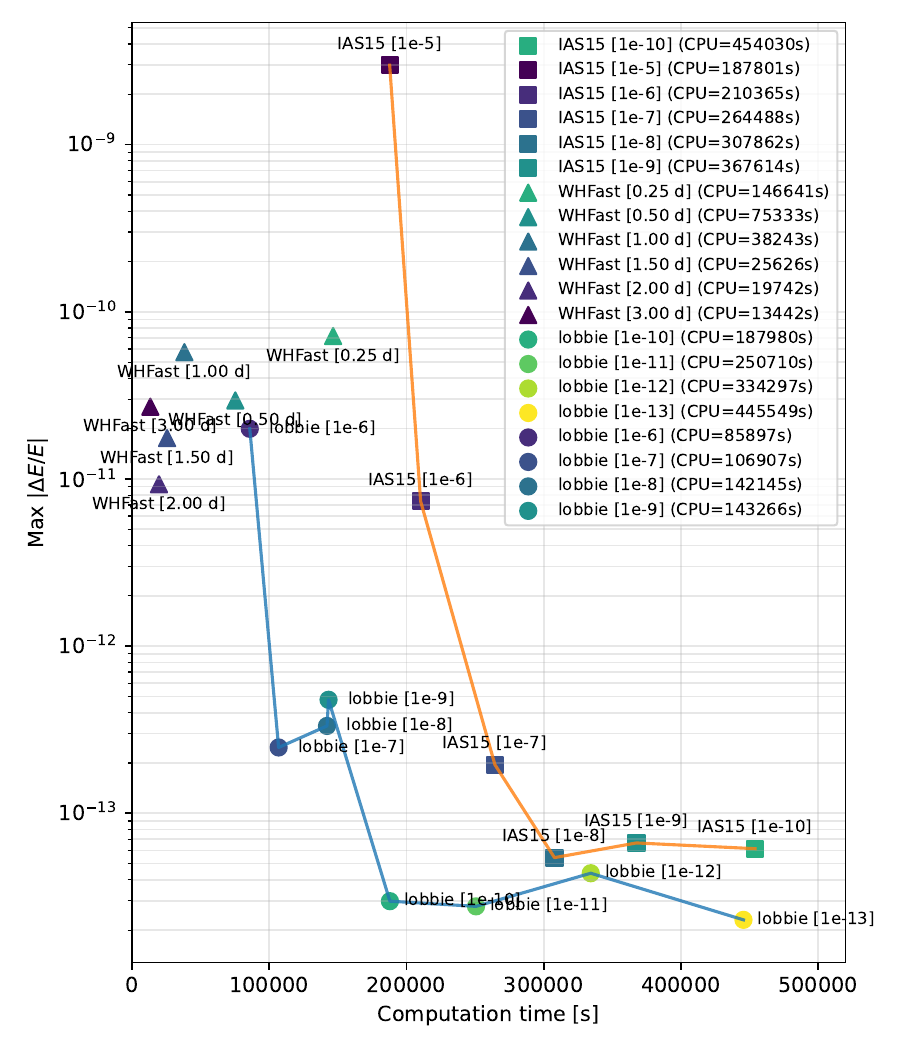} 
    \caption{The maximal relative error in total energy for different integrators vs computation time for the test of integrating 8 planets of the Solar system for 100~Myr. At comparable accuracy, Lobbie is about as fast as WHFast, but can be more accurate than IAS15 if required showing a significant performance boost compared to the latter.}
    \label{fig:8pl_max_e_vs_cpu}
\end{figure}

We first integrated all eight planets for 100 Myr. Since no analytical solution exists, we evaluated the performance by monitoring the conservation of total energy. The relative energy error was recorded every 250 years. The maximum relative energy error as a function of computation time is shown in Fig.~\ref{fig:8pl_max_e_vs_cpu}. For adaptive integrators (IAS15 and Lobbie), the timestep is mainly controlled by Mercury, which has the shortest orbital period.

Lobbie achieves very high accuracy. For $e_{tol}=10^{-13}$, the maximum relative energy error is $2.31 \cdot 10^{-14}$ with a computation time of 445,549 seconds (5.16 days). However, a good balance between accuracy and performance is already reached at $e_{tol}=10^{-10}$, which requires 187,980 seconds (2.16 days) and maintains the error below $3 \cdot 10^{-14}$, while being faster than all IAS15 runs. The best result for IAS15 is $5.5 \cdot 10^{-14}$ at $e_{tol}=10^{-8}$ requiring 307,862 seconds (3.56~days). WHFast is  faster, but its accuracy is significantly lower: for a timestep of 2 days, the error is $9.3 \cdot 10^{-12}$, more than two orders of magnitude worse.

\begin{figure}
    \centering
    \includegraphics[width=1\columnwidth]{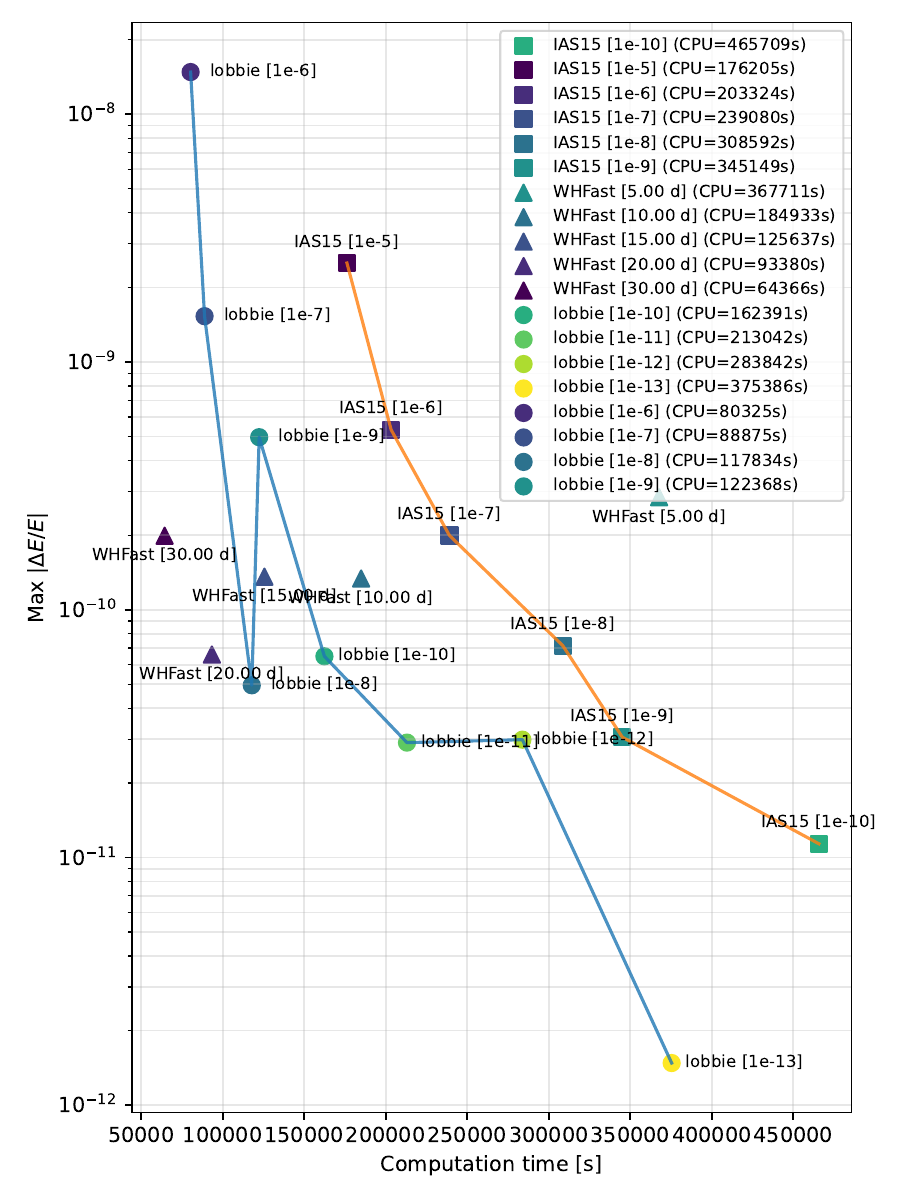} 
    \caption{Maximum relative energy error versus computation time for the integration of the four gas giants over 10 Gyr. At integration times comparable to the lifetime of the sun, symplectic integrators start to become competitive in terms of energy error, but again Lobbie can match its performance and can be tuned to either similar speed or higher accuracy.}
    \label{fig:4pl_max_e_vs_cpu}
\end{figure}

As an additional test, we integrated only the four gas giants for 10 Gyr. In this case, larger timesteps are possible because these planets are further from the Sun. The relative energy error was recorded every 10,000 years. For WHFast the stepsizes were also increased to $\{5, 10, 15, 20, 30 \}$ days. The results are shown in Fig.~\ref{fig:4pl_max_e_vs_cpu}.

Again, Lobbie provides the best accuracy. For $e_{tol}=10^{-13}$, the relative energy error is $1.48 \cdot 10^{-12}$, which is about an order of magnitude better than IAS15 with $e_{tol}=10^{-10}$, while also being faster. For $e_{tol}=10^{-11}$ and $10^{-12}$, Lobbie still achieves errors below $3 \cdot 10^{-11}$ with significantly shorter computation times. WHFast remains the fastest method, but with somewhat lower accuracy (error $\gtrsim 10^{-10}$), which is not easily remedied.

Overall, our results show that there is no universal choice of the accuracy parameter that provides the best performance for all problems. For Lobbie, the value $e_{tol}=10^{-11}$ appears to give a good balance between accuracy and speed across all tests. However, the optimal choice depends on the specific system.

This behavior can be understood physically. For Lobbie the parameter $e_{tol}$ controls the relative error in velocity per step. In the eight-planet case, the timestep is determined by Mercury, while the total energy is dominated by Jupiter. In contrast, in the four-planet case, Jupiter dominates both the timestep and the total energy. This explains why decreasing $e_{tol}$ improves the results for the four-planet system, but has no effect for the full Solar System, because it already reached the highest precision. A similar trend is observed for IAS15. WHFast does not show this behavior.

\subsection{Solar System with a population of asteroids}

\begin{figure}
    \centering
    \includegraphics[width=1\columnwidth]{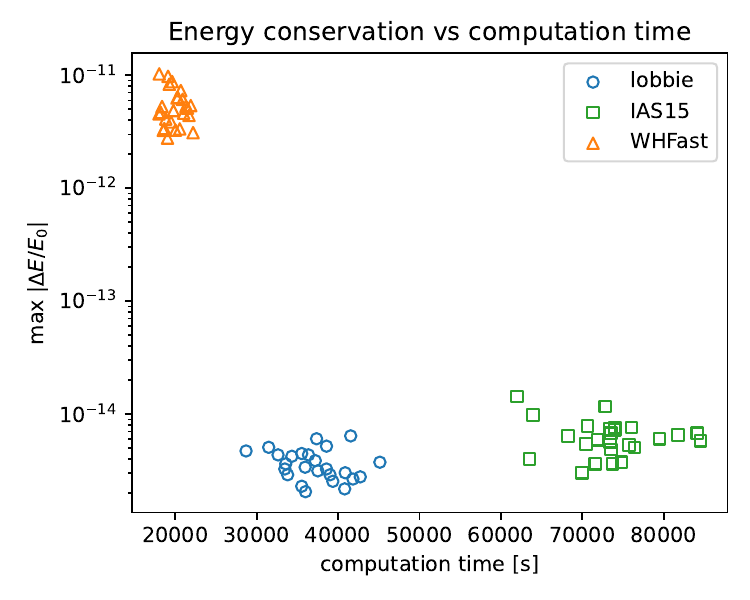} 
    \caption{\Dima{Maximum relative energy error versus average computation time, for 25 realizations each of Lobbie ($e_{tol}=10^{-11}$), IAS15 ($e_{tol}=10^{-8}$), and WHFast (timestep varied from 6~h in 1\% increments between runs), for the combined integration of eight planets and 50 Aten asteroids over 1~Myr. Lobbie and IAS15 exhibit smaller relative deviations from initial energy by about three orders of magnitude than WHFast, with Lobbie being roughly twice as fast as IAS15.}}
    \label{fig:planets_asteroids_energy}
\end{figure}

\begin{figure}
    \centering
    \includegraphics[width=1\columnwidth]{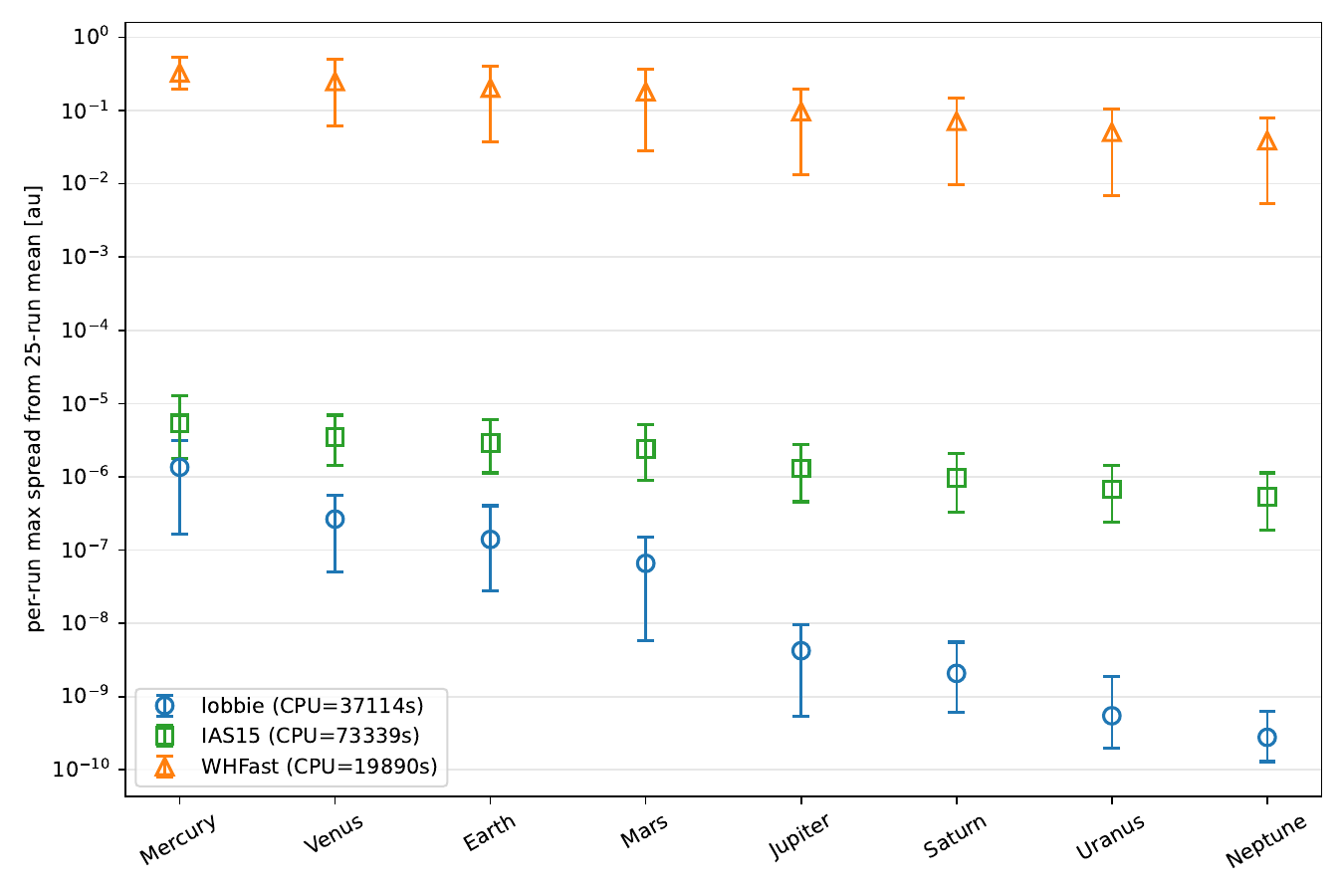} 
    \caption{\Dima{Positional spread across 25 runs of the same integrator, over the 1~Myr integration of eight planets and 50 Aten asteroids, shown separately for each planet. For each run, we compute the maximum deviation of that run's position from the mean position over all 25 runs; the markers show the average of this maximum deviation across the 25 runs, with whiskers spanning its minimum to maximum. Lobbie shows the smallest spread for every planet (by orders of magnitude), followed by IAS15, while WHFast is several orders of magnitude worse throughout, reflecting its much lower intrinsic accuracy.}}
    \label{fig:planets_asteroids_position}
\end{figure}

\begin{figure}
    \centering
    \includegraphics[width=1\columnwidth]{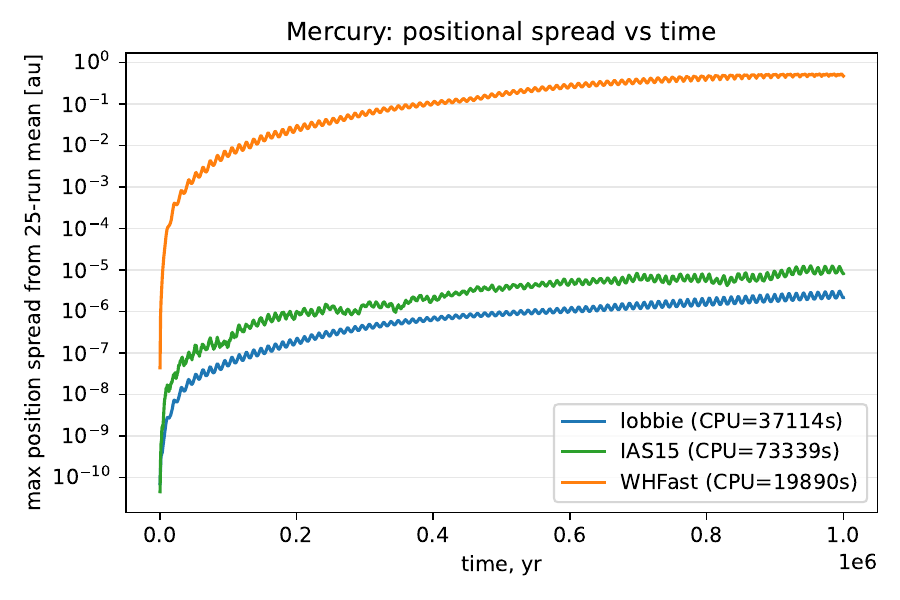} 
    \includegraphics[width=1\columnwidth]{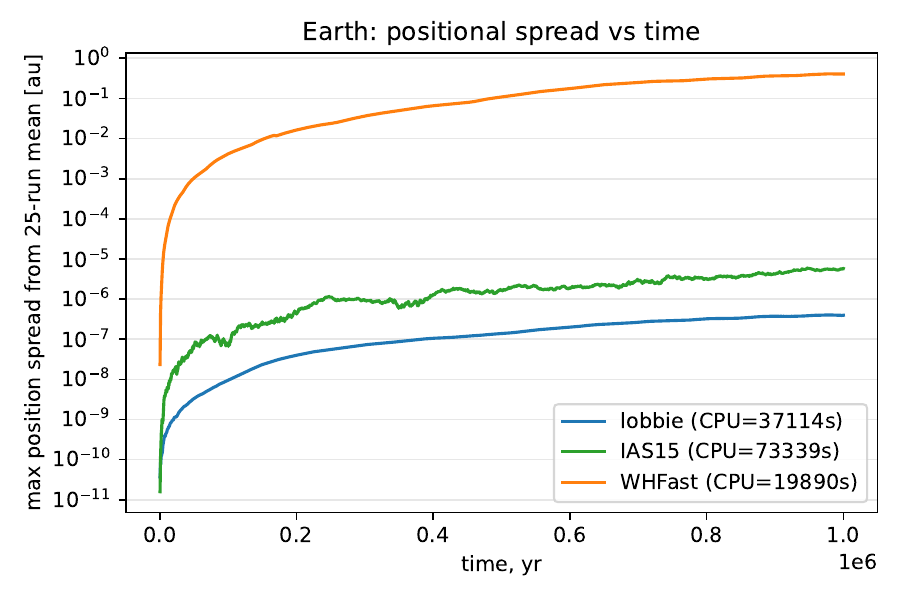} 
    \includegraphics[width=1\columnwidth]{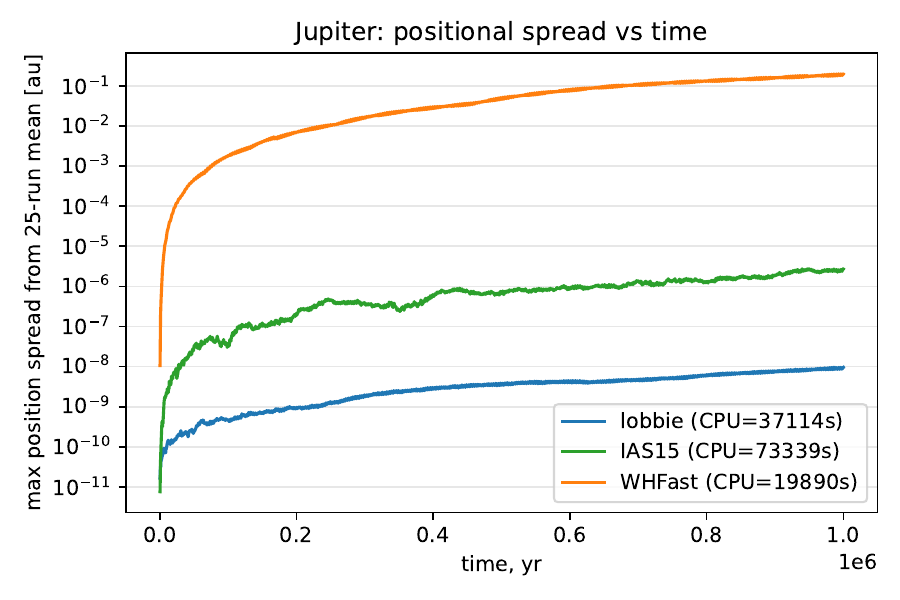} 
    \caption{\Dima{Maximum positional spread among 25 runs as a function of time, for Mercury, Earth and Jupiter (top to bottom), in the same eight-planet-plus-asteroids test as Figs.~\ref{fig:planets_asteroids_energy} and \ref{fig:planets_asteroids_position}. Lobbie and IAS15 grow gradually and stay close to round-off level, while WHFast diverges rapidly to $0.1-1$~au.}}
    \label{fig:planets_asteroids_planets_position}
\end{figure}

\Siggi{One of the challenges of numerical experiments including small body populations is consistency, i.e. how reproducible outcomes are. Close approaches between small bodies and the planets can lead to changes in the step-size of variable time-stepping integrators making integration outcomes dependent on the exact population and integration parameters used. To study the effects of variable timestepping on the consistency of planetary positions,}
\Dima{we combined the eight major planets with 50 massless Aten asteroids, whose orbits were taken at random from the NEOMOD3 model \citep{2024Icar..41716110N}. Aten asteroids cross the orbit of the Earth, so they regularly experience close encounters with the terrestrial planets. This makes the test considerably more chaotic than the previous ones, and, unlike the two-body case, no analytical solution is available to measure the true error directly.}

\Dima{To evaluate reliability in this chaotic regime, we instead used an ensemble approach: for each integrator we performed 25 independent realizations of the system and measured how much the individual runs spread apart from each other. For each of the 25 realizations, we drew a new random set of 50 Aten asteroids from the NEOMOD3 model, so that the runs differ from each other in their asteroid population rather than in the planets' initial conditions. For Lobbie we used $e_{tol}=10^{-11}$ and for IAS15 $e_{tol}=10^{-8}$, the values that performed best in the eight-planet test. Each realization is integrated with the same adaptive step size control, so any spread we measure in the planets' positions reflects only how sensitive the planetary orbits are to the presence of a different set of massless asteroids. WHFast, however, uses a fixed step and does not detect or respond to close encounters at all, so simply changing the asteroid population between runs would not change its behaviour in any meaningful way. To still be able to measure a spread for WHFast, we therefore also varied its timestep slightly between runs: the first realization used a step of 6~h, and each following realization increased the step by 1\%, so that the 25th run used a step about 24\% larger than the first.}

\Dima{Each system was integrated for 1~Myr. We recorded the positions of all bodies every 250~years and compared them across the 25 runs of each integrator, as well as monitoring the conservation of total energy.}

\Dima{The maximum relative energy error across all 25 runs is $6.40\cdot10^{-15}$ for Lobbie, $1.42\cdot10^{-14}$ for IAS15, and $9.75\cdot10^{-12}$ for WHFast, i.e. Lobbie  conserves the total energy about 2 times better, while being twice faster, and they both conserve energy roughly three orders of magnitude better than WHFast (Fig.~\ref{fig:planets_asteroids_energy}). In terms of computation time, WHFast is fastest at 19,890~s on average, Lobbie takes 37,114~s, and IAS15 is the slowest at 73,339~s.}

\Dima{Figure~\ref{fig:planets_asteroids_position} shows the maximum positional spread among the 25 runs at the end of the integration, for each planet and each integrator. Lobbie shows the smallest spread for every planet, typically one to two orders of magnitude smaller than IAS15, while WHFast is many orders of magnitude worse for all planets, including those far from the asteroid belt. For Jupiter, in particular, Lobbie's spread after 1~Myr remains below $10^{-8}$~au, i.e. about 1.5~km, despite the presence of 50 asteroids repeatedly perturbing the system. This level of consistency suggests that our realization of Lobbie is well suited to accurate long-term integrations where close encounters must be resolved reliably, such as the dynamical evolution of long-period comets during their encounters with Jupiter.}

\Dima{Lobbie also shows a clear downward trend, becoming more precise for planets further from the Sun, and this trend is much stronger than for IAS15. This is likely because the integration step is set mainly by the innermost planets and by close encounters with asteroids, making it far smaller than what the outer planets themselves would require; since Lobbie stores its state vectors in quadruple precision, it does not lose this extra accuracy to round-off, and the outer planets' positions come out correspondingly more precise. Figure~\ref{fig:planets_asteroids_planets_position}  (Mercury, Earth and Jupiter, respectively) show how this spread builds up over time: for Lobbie and IAS15 the spread grows gradually and stays close to the round-off level for most of the integration, whereas for WHFast it grows very rapidly at the start and quickly reaches $0.1-1$~au.}

\Dima{This last result must be interpreted with some care. Because WHFast does not detect close encounters, it integrates every asteroid with the same fixed step regardless of how close it comes to a planet, and it does so identically in every run except for the small artificial change in step size that we introduced by hand. The large spread we measure for WHFast is therefore driven mainly by this imposed step variation, not by an actual sensitivity of the code to the physical close encounters. In other words, this test somewhat favours WHFast: since it never adapts to the encounters in the first place, its apparent short computation time and its energy conservation do not reflect whether the close encounters were resolved correctly, only that the fixed step was small enough not to blow up the energy. For this reason, we do not consider the WHFast results in this asteroid test to be a reliable measure of its accuracy for systems with frequent close encounters, and the comparison here should be read mainly as a demonstration that Lobbie and IAS15, which do adapt their step size, remain self-consistent across many close encounters, while WHFast simply does not experience them in a numerically meaningful way.}

\section{Conclusions}

In this work, we evaluated the performance of the Lobatto collocation integrator Lobbie \citep{2022SoSyR..56...32A} for long-term gravitational $N$-body simulations of the Solar System and compared it against two widely used state-of-the-art integrators: the symplectic mapping WHFast and the adaptive high-order collocation integrator IAS15.  

Our tests show that Lobbie combines several desirable properties for long-term celestial mechanics simulations. In the circular two-body problem integrated over 100 Myr, Lobbie showed exceptional long-term phase stability, maintaining positional errors far below those reached by either IAS15 or WHFast. In particular, for suitable choices of the adaptive accuracy parameter, the orbital phase error remained bounded at levels corresponding to spatial deviations substantially smaller than planetary scales over the entire integration interval.
%\Siggi{In systems with orbital eccentricity up to $e=0.999$ IAS15 and Lobbie produce similar performance. Highly eccentric orbit configurations, however, see Lobbie dominate in terms of accuracy by about half an order of magnitude over IAS15 and WHFast.}
\Dima{In systems with orbital eccentricity up to $e=0.999$ IAS15 and WHFast perform better than for a circular orbit, however, Lobbie still obtains more accurate results. }

For integrations of the full eight-planet Solar System over 100 Myrs, Lobbie achieved relative energy conservation at the level of a few parts in $10^{-14}$ while being substantially faster than IAS15 at comparable accuracy. WHFast remained the fastest method overall, but at significantly lower energy accuracy (2 orders of magnitude worse). We furthermore demonstrate that Lobbie can approached WHFast-level performance while retaining the flexibility of a general-purpose adaptive collocation scheme capable of handling arbitrary force models and non-Hamiltonian perturbations.  
The integrations of the four giant planets over 10 Gyrs further demonstrated the long-term robustness of the method. On timescales comparable to the main-sequence lifetime of the Sun, Lobbie consistently produced smaller energy errors than IAS15 while maintaining competitive runtimes. The results indicate that symmetric Lobatto collocation schemes can exhibit long-term numerical behavior comparable to that traditionally associated with symplectic methods, despite not being explicitly symplectic themselves.  
\Siggi{As for consistency in systems that exhibit chaotic behavior, Lobbie again takes the lead over its competitors, especially in positional consistency of planets in the outer solar system} \Dima{achieving the order of 1~km agreement.}

% The present implementation also demonstrates that careful treatment of numerical precision can significantly improve long-term integrations. The use of quadruple-precision storage for state vectors reduced round-off accumulation at minimal computational overhead, while preserving the efficiency advantages of predominantly double-precision arithmetic.  
In summary, our results show that Lobbie represents a highly competitive alternative for long-term gravitational dynamics, \Dima{especially if high precision is required}. The method combines the geometric advantages of symmetric collocation schemes with adaptive timestepping, high-order accuracy, and broad applicability to general force models. These properties make Lobbie particularly attractive for applications involving close encounters, non-conservative perturbations, relativistic corrections, or long-term integrations of large populations of small bodies in the Solar System.

%%%%%%%%%%%%%%%%%%%%%%%%%%%%%%%%%%%%%%%%%%%%%%%%%%
\section*{Data Availability}

The code used in this work is publicly available. The Fortran implementation of the Lobbie integrator, including the version designed for integrations of massive bodies together with massless particles, can be accessed through the GitHub repository:
$\mathrm{https://github.com/DJVil/Lobbie\_planet\_integration.git}$

The data generated in this study are available from the corresponding author upon reasonable request.

\section*{Acknowledgements}
This research is supported by the National Science Foundation under grant Nos. 2307569 and 2307570 and is
partially funded by a generous gift of Charles Simonyi to the NSF Division of Astronomical Sciences. The award is made in recognition of significant contributions to Rubin Observatory’s Legacy Survey of Space and Time. Any opinions, findings, and conclusions or recommendations expressed in this material are those of the author(s) and do not necessarily reflect the views of the National Science Foundation.

The work of S.G. is supported by NOIRLab, which is managed by the Association of Universities for Research in Astronomy (AURA) under a cooperative agreement with the US National Science Foundation.



\bibliographystyle{mnras}
\bibliography{example} % if your bibtex file is called example.bib

@ARTICLE{2018MNRAS.475.5570H,
       author = {{Hernandez}, David M. and {Bertschinger}, Edmund},
        title = "{Time-symmetric integration in astrophysics}",
      journal = {\mnras},
         year = 2018,
        month = apr,
       volume = {475},
       number = {4},
        pages = {5570-5584},
          doi = {10.1093/mnras/sty184},
archivePrefix = {arXiv},
       eprint = {1708.07266},
 primaryClass = {astro-ph.IM},
       adsurl = {https://ui.adsabs.harvard.edu/abs/2018MNRAS.475.5570H}
}

@ARTICLE{2024Icar..41716110N,
       author = {{Nesvorn{\'y}}, David and {Vokrouhlick{\'y}}, David and {Shelly}, Frank and {Deienno}, Rogerio and {Bottke}, William F. and {Fuls}, Carson and {Jedicke}, Robert and {Naidu}, Shantanu and {Chesley}, Steven R. and {Chodas}, Paul W. and {Farnocchia}, Davide and {Delbo}, Marco},
        title = "{NEOMOD 3: The debiased size distribution of Near Earth Objects}",
      journal = {\icarus},
         year = 2024,
        month = jul,
       volume = {417},
          eid = {116110},
        pages = {116110},
          doi = {10.1016/j.icarus.2024.116110},
archivePrefix = {arXiv},
       eprint = {2404.18805},
 primaryClass = {astro-ph.EP},
       adsurl = {https://ui.adsabs.harvard.edu/abs/2024Icar..41716110N}
}

@ARTICLE{2023AJ....166...55N,
       author = {{Nesvorn{\'y}}, David and {Deienno}, Rogerio and {Bottke}, William F. and {Jedicke}, Robert and {Naidu}, Shantanu and {Chesley}, Steven R. and {Chodas}, Paul W. and {Granvik}, Mikael and {Vokrouhlick{\'y}}, David and {Bro{\v{z}}}, Miroslav and {Morbidelli}, Alessandro and {Christensen}, Eric and {Shelly}, Frank C. and {Bolin}, Bryce T.},
        title = "{NEOMOD: A New Orbital Distribution Model for Near-Earth Objects}",
      journal = {\aj},
         year = 2023,
        month = aug,
       volume = {166},
       number = {2},
          eid = {55},
        pages = {55},
          doi = {10.3847/1538-3881/ace040},
archivePrefix = {arXiv},
       eprint = {2306.09521},
 primaryClass = {astro-ph.EP},
       adsurl = {https://ui.adsabs.harvard.edu/abs/2023AJ....166...55N}
}

@ARTICLE{2018Icar..312..181G,
       author = {{Granvik}, Mikael and {Morbidelli}, Alessandro and {Jedicke}, Robert and {Bolin}, Bryce and {Bottke}, William F. and {Beshore}, Edward and {Vokrouhlick{\'y}}, David and {Nesvorn{\'y}}, David and {Michel}, Patrick},
        title = "{Debiased orbit and absolute-magnitude distributions for near-Earth objects}",
      journal = {\icarus},
         year = 2018,
        month = sep,
       volume = {312},
        pages = {181-207},
          doi = {10.1016/j.icarus.2018.04.018},
archivePrefix = {arXiv},
       eprint = {1804.10265},
 primaryClass = {astro-ph.EP},
       adsurl = {https://ui.adsabs.harvard.edu/abs/2018Icar..312..181G}
}

@ARTICLE{2012Icar..217..355G,
       author = {{Greenstreet}, Sarah and {Ngo}, Henry and {Gladman}, Brett},
        title = "{The orbital distribution of Near-Earth Objects inside Earth's orbit}",
      journal = {\icarus},
         year = 2012,
        month = jan,
       volume = {217},
       number = {1},
        pages = {355-366},
          doi = {10.1016/j.icarus.2011.11.010},
       adsurl = {https://ui.adsabs.harvard.edu/abs/2012Icar..217..355G}
}

@article{Duncan_1998,
doi = {10.1086/300541},
url = {https://doi.org/10.1086/300541},
year = {1998},
month = {oct},
publisher = {},
volume = {116},
number = {4},
pages = {2067},
author = {Duncan, Martin J. and Levison, Harold F. and Lee, Man Hoi},
title = {A Multiple Time Step Symplectic Algorithm for Integrating Close
Encounters},
journal = {The Astronomical Journal}
}

@incollection{eggl2009introduction,
  title={An introduction to common numerical integration codes used in dynamical astronomy},
  author={Eggl, Siegfried and Dvorak, Rudolph},
  booktitle={Dynamics of small solar system bodies and exoplanets},
  pages={431--480},
  year={2009},
  publisher={Springer}
}

@article{javaheri2023whfast512,
  title={WHFast512: A symplectic N-body integrator for planetary systems optimized with AVX512 instructions},
  author={Javaheri, Pejvak and Rein, Hanno and Tamayo, Dan},
  journal={The Open Journal of Astrophysics},
  volume={6},
  year={2023},
  publisher={Maynooth Academic Publishing}
}

@article{laskar2001high,
  title={High order symplectic integrators for perturbed Hamiltonian systems},
  author={Laskar, Jacques and Robutel, Philippe},
  journal={Celestial Mechanics and Dynamical Astronomy},
  volume={80},
  number={1},
  pages={39--62},
  year={2001},
  publisher={Springer}
}

@article{rein2015whfast,
  title={WHFAST: a fast and unbiased implementation of a symplectic Wisdom--Holman integrator for long-term gravitational simulations},
  author={Rein, Hanno and Tamayo, Daniel},
  journal={Monthly Notices of the Royal Astronomical Society},
  volume={452},
  number={1},
  pages={376--388},
  year={2015},
  publisher={Oxford University Press}
}

@article{rein2019high,
  title={High-order symplectic integrators for planetary dynamics and their implementation in rebound},
  author={Rein, Hanno and Tamayo, Daniel and Brown, Garett},
  journal={Monthly Notices of the Royal Astronomical Society},
  volume={489},
  number={4},
  pages={4632--4640},
  year={2019},
  publisher={Oxford University Press}
}

@article{rein2019symplectic,
    author = {Rein, Hanno and Brown, Garett and Tamayo, Daniel},
    title = {On the accuracy of symplectic integrators for secularly evolving planetary systems},
    journal = {Monthly Notices of the Royal Astronomical Society},
    volume = {490},
    number = {4},
    pages = {5122-5133},
    year = {2019},
    month = {12},
    issn = {0035-8711},
    doi = {10.1093/mnras/stz2942},
    url = {https://doi.org/10.1093/mnras/stz2942},
    eprint = {https://academic.oup.com/mnras/article-pdf/490/4/5122/30725851/stz2942.pdf},
}

@article{wisdom1996symplectic,
  title={Symplectic correctors},
  author={Wisdom, Jack and Holman, M and Touma, J},
  journal={Integration Algorithms and Classical Mechanics},
  volume={10},
  number={217-244},
  pages={180},
  year={1996},
  publisher={AMS, Providence}
}

@article{10.1145/363707.363723,
author = {Kahan, W.},
title = {Pracniques: further remarks on reducing truncation errors},
year = {1965},
issue_date = {Jan. 1965},
publisher = {Association for Computing Machinery},
address = {New York, NY, USA},
volume = {8},
number = {1},
issn = {0001-0782},
url = {https://doi.org/10.1145/363707.363723},
doi = {10.1145/363707.363723},
journal = {Commun. ACM},
month = jan,
pages = {40},
numpages = {2}
}

@ARTICLE{2021AJ....161..105P,
       author = {{Park}, Ryan S. and {Folkner}, William M. and {Williams}, James G. and {Boggs}, Dale H.},
        title = "{The JPL Planetary and Lunar Ephemerides DE440 and DE441}",
      journal = {\aj},
         year = 2021,
        month = mar,
       volume = {161},
       number = {3},
          eid = {105},
        pages = {105},
          doi = {10.3847/1538-3881/abd414},
       adsurl = {https://ui.adsabs.harvard.edu/abs/2021AJ....161..105P}
}

@article{Avdyushev2021,
  author    = {Avdyushev, V. A.},
  title     = {New Collocation Integrator for Solving Dynamic Problems.
               {I}. Theoretical Background},
  journal   = {Russian Physics Journal},
  year      = {2021},
  volume    = {63},
  number    = {11},
  pages     = {1977--1988},
  doi       = {10.1007/s11182-021-02260-2},
  language  = {english}
}

@article{Avdyushev2020,
  author    = {Avdyushev, V. A.},
  title     = {New Collocation Integrator for Solving Dynamic Problems.
               {I}. Theoretical Background},
  journal   = {Izvestiya Vysshikh Uchebnykh Zavedenii. Fizika},
  year      = {2020},
  volume    = {63},
  number    = {11},
  pages     = {131--140},
  doi       = {10.17223/00213411/63/11/131},
  language  = {russian}
}

@ARTICLE{2022SoSyR..56...32A,
       author = {{Avdyushev}, V.~A.},
        title = "{Collocation Integrator Lobbie in Orbital Dynamics Problems}",
      journal = {Solar System Research},
         year = 2022,
        month = jan,
       volume = {56},
       number = {1},
        pages = {32-42},
          doi = {10.1134/S0038094622010014},
       adsurl = {https://ui.adsabs.harvard.edu/abs/2022SoSyR..56...32A}
}

@book{hairer2006geometric,
  author    = {Hairer, Ernst and Lubich, Christian and Wanner, Gerhard},
  title     = {Geometric Numerical Integration: Structure-Preserving Algorithms for Ordinary Differential Equations},
  series    = {Springer Series in Computational Mathematics},
  volume    = {31},
  edition   = {2},
  year      = {2006},
  publisher = {Springer-Verlag},
  address   = {Berlin, Heidelberg},
  isbn      = {978-3-540-30663-4}
}

@ARTICLE{1974CeMec..10...35E,
       author = {{Everhart}, Edgar},
        title = "{Implicit Single-Sequence Methods for Integrating Orbits}",
      journal = {Celestial Mechanics},
         year = 1974,
        month = aug,
       volume = {10},
       number = {1},
        pages = {35-55},
          doi = {10.1007/BF01261877},
       adsurl = {https://ui.adsabs.harvard.edu/abs/1974CeMec..10...35E}
}

@ARTICLE{2015MNRAS.446.1424R,
       author = {{Rein}, Hanno and {Spiegel}, David S.},
        title = "{IAS15: a fast, adaptive, high-order integrator for gravitational dynamics, accurate to machine precision over a billion orbits}",
      journal = {\mnras},
         year = 2015,
        month = jan,
       volume = {446},
       number = {2},
        pages = {1424-1437},
          doi = {10.1093/mnras/stu2164},
archivePrefix = {arXiv},
       eprint = {1409.4779},
 primaryClass = {astro-ph.EP},
       adsurl = {https://ui.adsabs.harvard.edu/abs/2015MNRAS.446.1424R}
}

@Inbook{Feng2010,
author="Feng, Kang
and Qin, Mengzhao",
title="Symplectic Difference Schemes for Hamiltonian Systems",
bookTitle="Symplectic Geometric Algorithms for Hamiltonian Systems",
year="2010",
publisher="Springer Berlin Heidelberg",
address="Berlin, Heidelberg",
pages="187--211",
isbn="978-3-642-01777-3",
doi="10.1007/978-3-642-01777-3_5",
url="https://doi.org/10.1007/978-3-642-01777-3_5"
}

@ARTICLE{1991AJ....102.1528W,
       author = {{Wisdom}, Jack and {Holman}, Matthew},
        title = "{Symplectic maps for the N-body problem.}",
      journal = {\aj},
         year = 1991,
        month = oct,
       volume = {102},
        pages = {1528-1538},
          doi = {10.1086/115978},
       adsurl = {https://ui.adsabs.harvard.edu/abs/1991AJ....102.1528W}
}

@ARTICLE{1983ITNS...30.2669R,
       author = {{Ruth}, R.~D.},
        title = "{A Canonical Integration Technique}",
      journal = {IEEE Transactions on Nuclear Science},
         year = 1983,
        month = aug,
       number = {4},
        pages = {2669},
          doi = {10.1109/TNS.1983.4332919},
       adsurl = {https://ui.adsabs.harvard.edu/abs/1983ITNS...30.2669R}
}

@article{Vogelaere1900,
author={Vogelaere, Rene De},
title={Methods of Integration which Preserve the Contact Transformation Property of the Hamilton Equations (Version 1)},
journal={University of Notre Dame},
year=1900,
}

%\begin{thebibliography}{99}
%\end{thebibliography}

%%%%%%%%%%%%%%%%%%%%%%%%%%%%%%%%%%%%%%%%%%%%%%%%%%

%%%%%%%%%%%%%%%%% APPENDICES %%%%%%%%%%%%%%%%%%%%%

\appendix

\section{Full Lobbie algorithm}
\label{app:lobbie_full}

%Here we present the direct scheme to obtain  the state vector solutions for integrator Lobbie. The state vector solution is approximated by vector polynomials: $\bm{u}(t) \approx \bm{x}(t)$ and $\bm{v}(t) = \bm{\dot{u}}(t) \approx \bm{\dot{x}}(t)$. As we mentioned above the polynomials must satisfy Equation~(\ref{eq:diff_equation}) at collocation times $\{t_i\}_{i=1}^{s}$. For Lobbie we use Lobatto collocation nodes, therefore $t_1 = t_0$ and $t_s = t_0 + h$.  

In this appendix we provide the complete formulation of the collocation integrator \textsc{Lobbie}, following the original derivation of \citet{Avdyushev2021}. The method is formulated here only for second-order differential equations, consistent with the implementation used in this work.

\subsection{Collocation formulation}

Consider the second-order system
\begin{equation}
\ddot{\bm{x}} = \bm{f}(t,\bm{x},\dot{\bm{x}}),
\label{eq:lobbie_second_order}
\end{equation}
with initial conditions
\begin{align}
\bm{x}(t_0) &= \bm{x}_0,
\\
\dot{\bm{x}}(t_0) &= \bm{v}_0.
\end{align}

The integration interval is $[t_0,t_0+h]$, where $h$ is the integration step. Introducing the dimensionless variable
\begin{equation}
\tau = \frac{t-t_0}{h}, \qquad \tau \in [0,1],
\end{equation}
we define the Lobatto collocation nodes
\begin{equation}
0 = c_1 < c_2 < \dots < c_s = 1.
\end{equation}

The acceleration is approximated by the Newton interpolation polynomial
\begin{equation}
\bm{p}(\tau)
=
\sum_{j=1}^{s}
\bm{\alpha}_j
\prod_{k=1}^{j-1}(\tau-c_k),
\label{eq:newton_polynomial}
\end{equation}
where $\bm{\alpha}_j$ are the divided differences corresponding to the acceleration function.

The divided differences are computed recursively:
\begin{align}
\bm{\alpha}_1 &= \bm{f}_1,
\\
\bm{\alpha}_j
&=
\frac{\bm{\alpha}_{j-1:j}-\bm{\alpha}_{j-1}}
{c_j-c_{j-k}},
\end{align}
where
\begin{align}
\bm{f}_i &= \bm{f}(t_i,\bm{u}_i,\bm{v}_i),
\\
 t_i &= t_0 + hc_i.
\end{align}

For the 14th-order Lobatto collocation scheme with \(s=8\) collocation
points, the shifted Gauss--Lobatto nodes on the interval \([0,1]\) are

\begin{equation}
\begin{aligned}
c_1 &= 0,\\
c_2 &= 0.0641299257451967,\\
c_3 &= 0.204149909283429,\\
c_4 &= 0.395350391048761,\\
c_5 &= 0.604649608951239,\\
c_6 &= 0.795850090716571,\\
c_7 &= 0.935870074254803,\\
c_8 &= 1.
\end{aligned}
\end{equation}
These nodes are obtained from the roots of $\frac{d}{dx}P_7(x)=0$, mapped from the interval \([-1,1]\) to \([0,1]\).

\subsection{Collocation equations}

The approximate solutions at the collocation nodes are obtained by integrating the interpolation polynomial. For each collocation point $i=2,\dots,s$,
\begin{align}
\bm{u}_i
&=
\bm{x}_0
+
 hc_i\bm{v}_0
+
 h^2
\sum_{j=1}^{s}
 a_{ij}
\bm{\alpha}_j,
\\
\bm{v}_i
&=
\bm{v}_0
+
 h
\sum_{j=1}^{s}
 b_{ij}
\bm{\alpha}_j,
\label{eq:lobbie_collocation}
\end{align}
where $\bm{u}_i$ and $\bm{v}_i$ are approximations of the coordinates and velocities at the collocation point $c_i$.

The coefficients are defined as integrals of the Newton basis functions,
\begin{align}
 a_{ij}
 &=
 \int_0^{c_i}
 \int_0^{\tau}
 \prod_{k=1}^{j-1}(\xi-c_k)
 \,d\xi\,d\tau,
 \\
 b_{ij}
 &=
 \int_0^{c_i}
 \prod_{k=1}^{j-1}(\tau-c_k)
 \,d\tau.
\label{eq:ab_coefficients}
\end{align}

After convergence of the iterative process, the solution at the end of the step is obtained from
\begin{align}
\bm{x}_1
&=
\bm{x}_0
+
 h\bm{v}_0
+
 h^2
\sum_{j=1}^{s}
 a_{sj}
\bm{\alpha}_j,
\\
\dot{\bm{x}}_1
&=
\bm{v}_0
+
 h
\sum_{j=1}^{s}
 b_{sj}
\bm{\alpha}_j.
\label{eq:lobbie_final_step}
\end{align}

\subsection{Recursive construction of the coefficients}

The coefficients of the method are efficiently computed through recursive relations. Defining the $k$-fold integral of the Newton basis functions,
\begin{equation}
\gamma_j^{(k)}(\tau)
=
\underbrace{\int_0^{\tau}\int_0^{\tau_1}\dots\int_0^{\tau_{k-1}}}_{k\text{ times}}
\prod_{m=1}^{j-1}(\tau_k-c_m)
\,d\tau_k\dots d\tau_1,
\end{equation}
we have
\begin{align}
 a_{ij} &= \gamma_j^{(2)}(c_i), \\
 b_{ij} &= \gamma_j^{(1)}(c_i).
\end{align}

The recurrence relations are
\begin{align}
\gamma_1^{(k)}(\tau)
&=
\frac{\tau^k}{k!},
\\
\gamma_j^{(k)}(\tau)
&=
(\tau-c_{j-1})\gamma_{j-1}^{(k)}(\tau)
-
 k\gamma_{j-1}^{(k+1)}(\tau),
\label{eq:gamma_recurrence}
\end{align}
with $j=2,\dots,s$.

\subsection{Predictor--corrector iterations}

The nonlinear collocation equations are solved iteratively using a fixed-point iteration in Seidel form. Initial estimates of the divided differences are obtained by extrapolation from the previous step,
\begin{equation}
\bm{f}_i^{(0)} = \bm{p}(1+c_i).
\label{eq:predictor}
\end{equation}

At each iteration the collocation values are updated sequentially:
\begin{enumerate}
    \item Compute $\bm{u}_i$ and $\bm{v}_i$ from Eq.~(\ref{eq:lobbie_collocation}).
    \item Evaluate the accelerations
    \begin{equation}
    \bm{f}_i = \bm{f}(t_i,\bm{u}_i,\bm{v}_i).
    \end{equation}
    \item Recompute the divided differences $\bm{\alpha}_i$.
\end{enumerate}

The iterative process stops when
\begin{equation}
\|\bm{u}_s^{(n)}-\bm{u}_s^{(n-1)}\| < \varepsilon,
\label{eq:convergence_condition}
\end{equation}
where $\varepsilon$ is the convergence tolerance and $^{(n)}$ indicates the $n$-th iteration.

\subsection{Dense output}

One advantage of collocation methods is the possibility to construct a continuous approximation inside the integration step. For arbitrary $\tau \in [0,1]$,
\begin{align}
\bm{u}(\tau)
&=
\bm{x}_0
+
 h\tau\bm{v}_0
+
 h^2
\sum_{j=1}^{s}
\gamma_j^{(2)}(\tau)
\bm{\alpha}_j,
\\
\bm{v}(\tau)
&=
\bm{v}_0
+
 h
\sum_{j=1}^{s}
\gamma_j^{(1)}(\tau)
\bm{\alpha}_j.
\label{eq:dense_output}
\end{align}

This representation allows efficient generation of dense output without reintegration.

\subsection{Order and geometric properties}

For Lobatto collocation nodes, the integrator has formal order
\begin{equation}
 p = 2s-2,
\end{equation}
where $s$ is the number of collocation points. In addition, the method is symmetric, which makes it particularly suitable for long-term integrations in celestial mechanics.

The practical implementation of \textsc{Lobbie} follows the standard predictor--corrector structure:
\begin{enumerate}
    \item initialization of collocation coefficients,
    \item prediction of divided differences,
    \item iterative correction of collocation values,
    \item convergence check,
    \item dense output construction.
\end{enumerate}

\section{State vectors of the major planets}
\label{sec:state_vectors_of_planets}

Here in Table~\ref{tab:planets} we present the state vectors of 8 major planets we used in this computations and in Table~\ref{tab:masses} we present the masses of the planets.

\begin{table}
    \centering
    \begin{tabular}{c|c}
		Planet & Mass ($M_{\sun}$) \\ \hline
        Mercury & 1.660120825478540E-007 \\
		Venus & 2.447838287781388E-006\\
		Earth & 3.003489615446052E-006\\
		Mars & 3.227156082893391E-007 \\ 
		Jupiter & 9.547919099353572E-004 \\
		Saturn & 2.858856700227778E-004\\
		Uranus & 4.366249613194372E-005 \\
		Neptune & 5.151383772621838E-005\\
    \end{tabular}
    \caption{Mass of the planets used in our simulations in Solar Masses.}
    \label{tab:masses}
\end{table}

\begin{table*}
\centering

\label{tab:planets}

\caption{The initial heliocentric state vectors for all 8 major planets used in our tests. The distances are in au and the velocities are in au/days.}

\begin{tabular}{c  c}
\hline
Planet & State Vector $(x,y,z,v_x,v_y,v_z)$ \\
\hline

1 & \begin{tabular}[t]{r r r}
-0.140728079719301 & -0.443900958053739 & -2.334555920112938E-002 \\
 2.116887136020025E-002 & -7.097975420984262E-003 & -2.522831030870965E-003
\end{tabular} \\[2mm]

2 & \begin{tabular}[t]{r r r}
-0.718630216963728 & -2.250380069564102E-002 & 4.117184128884459E-002 \\
 5.135327471763992E-004 & -2.030614162369829E-002 & -3.071745200872245E-004
\end{tabular} \\[2mm]

3 & \begin{tabular}[t]{r r r}
-0.168524648395985 & 0.968783304965269 & -4.120490250836273E-006 \\
-1.723394583350394E-002 & -3.007660250034951E-003 & 3.562572767584586E-008
\end{tabular} \\[2mm]

4 & \begin{tabular}[t]{r r r}
1.39036106612293 & -2.100972225930386E-002 & -3.461801441143633E-002 \\
7.479271206652266E-004 & 1.518629868832040E-002 & 2.997532185989235E-004
\end{tabular} \\[2mm]

5 & \begin{tabular}[t]{r r r}
4.00346010820115 & 2.93535361601416 & -0.101823035649513 \\
-4.563473318088745E-003 & 6.446757879841354E-003 & 7.545651030030440E-005
\end{tabular} \\[2mm]

6 & \begin{tabular}[t]{r r r}
6.40855472782918 & 6.56804414319537 & -0.369127882426688 \\
-4.291119708408092E-003 & 3.891580683171841E-003 & 1.028767887671777E-004
\end{tabular} \\[2mm]

7 & \begin{tabular}[t]{r r r}
14.4305174938589 & -13.7356577542398 & -0.238129300641418 \\
2.678379969541383E-003 & 2.672442671258471E-003 & -2.477662656650322E-005
\end{tabular} \\[2mm]

8 & \begin{tabular}[t]{r r r}
16.8107573081082 & -24.9926513083273 & 0.127270834587742 \\
2.579369564538994E-003 & 1.776769336668958E-003 & -9.590902716799120E-005
\end{tabular} \\

\hline
\end{tabular}

\end{table*}

%%%%%%%%%%%%%%%%%%%%%%%%%%%%%%%%%%%%%%%%%%%%%%%%%%

% Don't change these lines
\bsp	% typesetting comment
\label{lastpage}
\end{document}